 \documentclass[twocolumn]{aastex62}

\shortauthors{Kato et al.}

\begin{document}
% not in full CAPITAL  -> Capital

\title{Production of Silicon on Mass Increasing White Dwarfs 
-- Possible Origin of High-Velocity-Features in Type Ia Supernovae}

\author{Mariko Kato} 
\affil{Department of Astronomy, Keio University, Hiyoshi, Yokohama
  223-8521, Japan;}
\email{mariko.kato@hc.st.keio.ac.jp}

\author{Hideyuki Saio}
\affil{Astronomical Institute, Graduate School of Science,
    Tohoku University, Sendai, 980-8578, Japan}
% \email{saio@astr.tohoku.ac.jp}
%\and

\author{Izumi Hachisu}
\affil{Department of Earth Science and Astronomy, College of Arts and
Sciences, The University of Tokyo, 3-8-1 Komaba, Meguro-ku,
Tokyo 153-8902, Japan}
%\email{hachisu@ea.c.u-tokyo.ac.jp}

\begin{abstract} 
Type Ia supernovae (SNe Ia) often show high-velocity absorption
features (HVFs) in their early phase spectra; however the origin of the HVFs is unknown. 
We show that a near-Chandrasekhar-mass white dwarf (WD) develops
a silicon-rich layer on top of a carbon-oxygen (CO) core 
before it explodes as an SN Ia.  
We calculated the nuclear yields in successive helium shell flashes  
for 1.0 $M_\sun$, 1.2 $M_\sun$, and 1.35 $M_\sun$ CO WDs accreting 
helium-rich matter with several mass-accretion rates ranging 
from $1 \times 10^{-7}~M_\sun$~yr$^{-1}$
to $7.5 \times 10^{-7}~M_\sun$~yr$^{-1}$.  
For the $1.35~M_\sun$ WD with the accretion rate of 
$1.6 \times 10^{-7}~M_\sun$~yr$^{-1}$, the surface layer 
developed as helium burning ash and 
consisted of 40\% $^{24}$Mg, 33\% $^{12}$C, 23\% $^{28}$Si, 
and a few percent of $^{20}$Ne by weight. 
For a higher mass accretion rate of $7.5 \times 10^{-7}~M_\sun$~yr$^{-1}$, 
the surface layer consisted of 58\% $^{12}$C, 31\% $^{24}$Mg, 
and 0.43\% $^{28}$Si.  For the $1.2~M_\sun$ WDs, 
silicon is produced only for lower mass accretion 
rates (2\% for $1.6 \times 10^{-7}~M_\sun$~yr$^{-1}$). 
No substantial silicon ($< 0.07\%$) is produced on the $1.0~M_\sun$ WD 
independently of the mass-accretion rate. 
If the silicon-rich surface layer is the origin of \ion{Si}{2} HVFs,
its characteristics are consistent with that of mass increasing WDs. 
We also discuss possible Ca production on very massive WDs 
($ \gtrsim 1.38~M_\sun$). 
\end{abstract}

% up to 6 keywords

\section{Introduction} \label{sec_introduction}

Type Ia supernovae (SNe Ia) are thought to be thermonuclear explosions 
of a carbon-oxygen (CO) white dwarf (WD) in a close binary system. 
However, the exact nature of the progenitor binary system and the details
of the explosion mechanism are still being debated. 

Many SNe Ia show high velocity absorption lines 
of \ion{Ca}{2} near-infrared (NIR)
triplet and \ion{Si}{2} $\lambda6355$ 
\citep{maz05,chi14,zha15,zha16} before or near $B$ maximum. 
Their line velocities are faster by $\sim6000$~km~s$^{-1}$ 
or more than the photospheric velocity; hence, these high-velocity
lines are referred to as high-velocity features (HVFs). 
Several ideas on the origin of HVFs have been proposed.
These ideas are associated with one of 
(1) Ca and Si abundance enhancements in the outermost layer of the ejecta, 
(2) density enhancement caused by swept-up material, 
and (3) ionization enhancement in the outermost layer of the ejecta
\citep[see, e.g.,][]{ger04, maz05, tan06}. 
This is because the outermost layers of SNe~Ia ejecta have the fastest
expansion velocities \citep[e.g., Figure 4 of][]{ger04}, the velocities of
which are much faster than the bulk of intermediate elements near 
the photosphere.

The origin of HVFs provides the nuclear burning process 
of SNe~Ia, especially in the outermost layers of the ejecta, and elucidates
the exact nature of their immediate progenitors.
In the present paper, we examine a possible origin of the aforementioned
abundance enhancement (1), that is, the nuclear burning yields in helium
shell flashes on massive CO WDs as proposed by \citet{kam12} and present 
the chemical composition of the outermost layer of progenitor WDs.
If all the mass-increasing WDs that will explode as an SN~Ia have 
developed silicon-rich layer on top of the CO core, it may explain
the nature of \ion{Si}{2} $\lambda6355$ HVFs.

Two major progenitor scenarios of SNe~Ia have been proposed so far 
\citep[e.g.,][]{mao14}.
One is the single-degenerate (SD) model, for which the binary system consists of 
a WD and a non-degenerate star like the main-sequence star,
and the other is the double-degenerate (DD) model, which consists
of two WDs.
There are many immediate progenitor models of the SD scenario
presented so far, including a WD with steady hydrogen burning, 
a recurrent nova (unstable hydrogen burning), 
and helium nova that accretes helium-rich material from a hydrogen-deficient
companion. In the former two cases, hydrogen burning produces 
helium ash underneath the hydrogen burning zone, and a helium flash 
occurs when the helium layer satisfies the ignition condition. 
Thus, many accreting WDs experience helium shell flashes 
and  nuclei heavier than carbon are inevitably yielded 
and accumulate on the CO core.  
The aim of this paper is to study helium nuclear burning products
for various WD masses and mass accretion rates, 
and to show that mass-accreting WDs are a possible production site of 
Si, which could be the origin of \ion{Si}{2} $\lambda6355$ HVFs. 

Herein, we concentrate on helium-rich matter accretion, because  
in the presence of hydrogen, a few hundred H-flashes should occur 
between two consecutive He-flashes, which causes our calculations 
to be extremely time-consuming. Instead, we assume that the WD
accretes hydrogen-deficient matter.  This paper is organized as follows;
we first briefly describe our numerical method 
in Section \ref{section_method}.
Our numerical results are presented in Section \ref{section_results}. 
In Section \ref{section_production_site}, we propose a production
site for \ion{Si}{2} $\lambda 6355$ HVFs.
In Section \ref{section_he_novae}, we present nuclear yield results of
helium novae and chemical composition of the ejecta.
The discussion and conclusions follow in Sections
\ref{section_discussion} and \ref{section_conclusion}, respectively.

%Fig.1
%\placefigure{HenomotoD} 
\begin{figure}
\epsscale{1.15}
% \plotone{iso_ignitime_he.eps}
%\plotone{dMwdmdot.He.ps}
\plotone{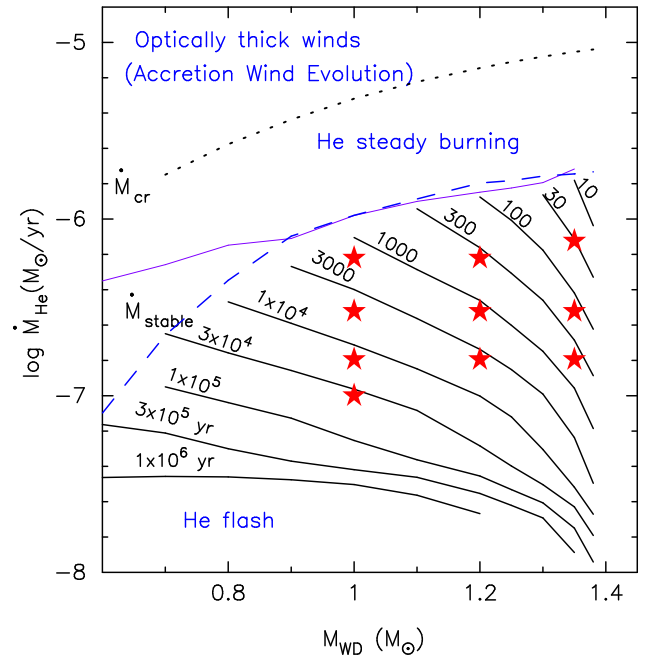}
\caption{
The response of WDs on the mass accretion rate.  
The chemical composition of the accreted matter is $Y=0.98$ and $Z=0.02$.  
Helium flashes occur below the stability line of He-burning 
($ \dot M_{\rm stable}$: dashed line).  The loci of equi-recurrence
period are plotted together with its recurrence period. 
Between the two lines of $ \dot M_{\rm cr}$ and $ \dot M_{\rm stable}$,
the He nuclear burning rate is the same as the mass-accretion rate.  
In the region above the dotted line 
($ \dot M_{\rm cr}$), the optically thick winds are accelerated 
and the binary undergoes accretion wind evolution 
\citep{hkn96, hkn99, hknu99}. 
The red stars indicate the models in Table \ref{table_model}. 
We added the stability line obtained by \citet{wan17} (thin solid line). 
See Section \ref{section_stability} for more details. 
\label{HenomotoD}}
% source: rnhe.mgsi/dMwdmdot.wip
\end{figure}

\section{Numerical Method}
\label{section_method}
 
We have calculated successive He shell flashes. The details of 
our numerical code and calculation method are published
in \citet{kat17sha, kat17shb}. 
Nucleosynthesis in the helium burning is calculated up to $^{28}$Si, in 
which nuclear  reaction rates and Q-values are obtained from 
\citet{cau88}, screening factors are obtained from \citet{gra73}. 
Neutrino emissions not related to the nuclear reactions are 
obtained from the formulae of \citet{ito89}. Coulomb effects in the 
equation of state in the dense core are included using the empirical 
equations obtained by \citet{van71}. 

The WD masses and mass accretion rates are summarized 
in Table \ref{table_model} and also indicated by 
the filled red stars in Figure \ref{HenomotoD}. 
Multicycle nova outbursts were calculated using a Henyey-type evolution
code.  The code encounters numerical difficulties when the nova envelope
expands to a giant size.  To continue numerical calculations beyond
the extended stage, we adopted numerical mass loss schemes because
during the extended stages of nova outbursts,
optically-thick wind mass-loss occurs \citep[e.g.,][]{kat94h}.
To avoid the complicated iteration process of wind fitting 
(that gives the proper wind mass-loss rate), we simply assume 
a mass-loss formula during the extended stage 
\citep{kat17sha}. 

We obtained mass loss rates as 
\begin{equation}
\dot{M}^{n+1}=\dot{M}^n(R^n_{\rm ph}/R^{n-1}_{\rm ph})^a,
\label{numerical-mass-loss}
\end{equation}
where $\dot{M}^{n+1}$ is the numerical mass-loss rate
and $R^{n+1}_{\rm ph}$ the photospheric radius for $(n+1)$ time-step. 
To calculate mass-loss stages successfully we chose a parameter 
$a$ as 4, 3, and 2 for $M_{\rm WD}=1.0,~1.2$, and $1.35~M_\sun$, respectively.
The initial mass loss rates were chosen as $-10^{-7}$ and 
$-10^{-6}M_\sun$~yr$^{-1}$. 
We started the mass loss when the photospheric radius
expanded to the given value, $\log (R_{0}/R_\sun)= -1.1$ to $-1.3$, 
and its rate increases with time. The mass loss rate decreases as 
the radius becomes smaller. We stop the mass loss when the radius becomes
smaller than the given value, $\log (R_{1}/R_\sun)= -1.5$ to $-1.4$. 
The maximum mass-loss rate for each model 
is given in Table \ref{table_model}.

%Table 1
%\placetable{table_model}
\begin{deluxetable*}{lllllrlllll}
\tabletypesize{\scriptsize}
\tablecaption{Model parameters of He shell flashes}
\tablewidth{0pt}
\tablehead{
\colhead{Model}&
\colhead{$M_{\rm WD}$}&
\colhead{$\dot M_{\rm acc}$}&
\colhead{$\dot M_{\rm ML}^{\rm max}$}&
\colhead{Cycle}&
\colhead{$P_{\rm rec}$}&
\colhead{$M_{\rm acc}$}&
\colhead{Flash$^{\rm a}$}&
\colhead{$L_{\rm nuc}^{\rm max}$}&
\colhead{$\log T_{\rm max}^{\rm max}$}&
\colhead{$\log T^*_{\rm max}$}\\
\colhead{}& 
\colhead{($M_\sun$)}&
\colhead{($M_\sun$yr$^{-1}$)}&
\colhead{($M_\sun$yr$^{-1}$)}&
\colhead{}&
\colhead{(yr)}&
\colhead{($M_\sun$)}&
\colhead{(yr)}&
\colhead{($L_\sun$)}&
\colhead{(K)}&
\colhead{(K)}
}
\startdata
M10.6  & 1.0  & 6.0$\times 10^{-7}$ &$1.2\times 10^{-5}$& 41 & 1270 &$7.4\times 10^{-4}$ &165 &$2.5 \times 10^{7}$ &8.54&8.56\\
M10.3  & 1.0  & 3.0$\times 10^{-7}$ &$8.8\times10^{-5}$& 44 & 7330 &$2.2\times 10^{-3}$ &111 &$6.2 \times 10^{9}$ &8.66&8.72\\
M10.16 & 1.0  & 1.6$\times 10^{-7}$ &$4.4\times10^{-4}$& 25 &$31800$ &$5.1\times 10^{-3}$&101 &$2.6 \times 10^{11}$&8.74 &8.83\\
M10.16.T$^{\rm b}$&1.0&1.6$\times 10^{-7}$&$1.6\times 10^{-2}$&13& $35300$ & $5.6\times 10^{-3}$ &56.0 &$3.5 \times 10^{11}$&8.74&8.84\\
M10.1  & 1.0  & 1.0$\times 10^{-7}$ &$1.3\times10^{-3}$& 55 &$74200$ & $7.4\times 10^{-3}$ &89.0&$1.5 \times 10^{12}$&8.78 &8.89\\
M12.6  & 1.2  & 6.0$\times 10^{-7}$ &$1.8\times 10^{-5}$& 20 & 444  &$ 2.6\times 10^{-4}$  &27.3&$7.1 \times 10^{7}$&8.62&8.63\\ 
M12.3  & 1.2  & 3.0$\times 10^{-7}$ &$5.1\times 10^{-5}$& 21 & 1990 &$ 5.9\times 10^{-4}$  &25.0&$3.3 \times 10^{9}$&8.71&8.74\\ 
M12.16 & 1.2  & 1.6$\times 10^{-7}$ &$2.8\times 10^{-4}$& 27 & 9340 &$ 1.5\times 10^{-3}$  &21.4& $2.5 \times 10^{11}$&8.81 &8.85\\
M135.75 &1.35 & 7.5$\times 10^{-7}$ &$5.3\times 10^{-6}$& 14 & 34.1 &$ 2.5\times 10^{-5}$ &3.42& $5.0 \times 10^{6}$&8.66&8.64\\
M135.3  &1.35 & 3.0$\times 10^{-7}$ &$1.7\times 10^{-5}$& 29 & 203  &$ 6.0\times 10^{-5}$ & 4.29&$5.0 \times 10^8 $& 8.74 &8.73\\
M135.16& 1.35 & 1.6$\times 10^{-7}$ &$3.5\times 10^{-5}$& 15 & 743  &$ 1.2\times 10^{-4}$ & 4.90&$8.5 \times 10^9$ &8.82& 8.81 \\
\enddata
\tablenotetext{a}{The flash duration is defined by the period during which
$L_{\rm ph} > 10^4 L_\sun$.} 
\tablenotetext{b}{A test model with extremely large mass-loss rate.}
\label{table_model}
\end{deluxetable*}

Table \ref{table_model} summarizes our model parameters.  
The first column shows the model name, i.e., 
M10.6 means that the WD mass is 1.0 $M_\sun$ with the mass accretion rate
of $6 \times 10^{-7}~M_\sun$~yr$^{-1}$. 
Then, the table lists the WD mass, mass accretion rate, 
maximum mass-loss rate, number of the flash cycles we calculated,
the recurrence period which is defined as the time between 
the epochs of maximum nuclear luminosity, accreted mass, 
flash duration, maximum nuclear luminosity $L_{\rm nuc}^{\rm max}$, 
the maximum temperature $T_{\rm max}^{\rm max}$ at one cycle of shell flash 
(i.e., the maximum of $T_{\rm max}$), and the maximum temperature 
$T_{\rm max}^*$ of the plane-parallel atmosphere, which is 
explained later in Equation (\ref{equation.T}). 

We assume that the CO core is composed of 48\% of $^{12}$C, 50\% of $^{16}$O,
and 2\% of $^{24}$Mg by weight. This assumption does not 
affect our results, because the CO core material
hardly mixes with the upper part and the newly accreted matter
is burned into heavy elements and accumulates on the surface of CO core. 
Our nuclear network includes 14 isotopes, i.e., $^1$H, $^3$He,
$^4$He, $^{12}$C, $^{13}$C, $^{14}$N, $^{15}$N, $^{16}$O, $^{17}$O, 
$^{18}$O, $^{20}$Ne, $^{22}$Ne, $^{24}$Mg, 
and $^{28}$Si coupled by 21 reactions. 
This nuclear network is small enough to efficiently speed up
the time-consuming calculation of a number of shell flashes, 
but covers all the main paths of nuclear reactions for both energy
and nuclide productions.  
The resultant nuclei distribution is in good agreement 
with other calculations which include much larger networks
as will be discussed in Section \ref{discussion_nuclei}. 

Figure \ref{HenomotoD} shows the response of WDs 
for the He mass-accretion rate vs. WD mass, which is the so-called
Nomoto diagram \citep{nom82}.
Here, we assume that the accreting matter is helium-rich and
has the composition of helium $Y=0.98$ and heavy elements $Z=0.02$
by weight, where $Z$ includes the solar composition of elements heavier
than helium.  Under the stability line, helium shell-burning is unstable
and periodic helium nova outbursts occur.  Between the stability line 
($\dot M_{\rm stable}$) and the line denoted by $\dot M_{\rm cr}$, 
helium shell-burning is stable with the consumption rate being the same
as the mass-accretion rate.  Above the line of $\dot M_{\rm cr}$, 
optically thick winds are accelerated and the WD undergoes 
accretion wind evolution, in which the WD accretes matter from 
the accretion disk and, at the same time, the WD loses mass from 
the other direction in the wind 
\citep[see, e.g., Figure 1 of][]{hac01kb}.   
Our calculated models correspond to the WD masses and mass-accretion rates
indicated with the filled red stars. 
We discuss the stability line in Section \ref{section_stability}.  

Although we started our calculation close to the thermal equilibrium model
\citep{nom07, kat14shn}, the recurrence periods (red stars)
in Table \ref{table_model} are slightly longer than those 
(black lines) in Figure \ref{HenomotoD}, because our time-dependent 
shell flash models have not yet reached a limit cycle. 
For example, model M10.1 has $P_{\rm rec}=79500$ yr in the 23th cycle
but $P_{\rm rec}=74200$ yr in the 55th cycle. 
We stopped our calculation before $P_{\rm rec}$ reach the limit cycle
because it converges very slowly \citep{kat17shb}.

%Fig.2
%\placefigure{m10.qx4}
\begin{figure}
\epsscale{1.15}
\plotone{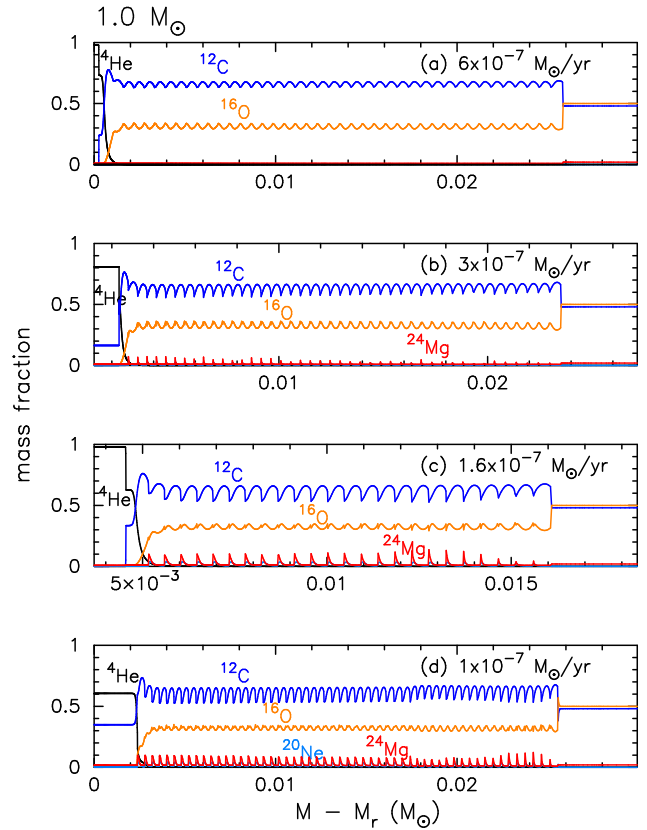}
%\plotone{m10qx.ps}
\caption{
Mass fraction of $^{12}$C (blue), $^{16}$O (orange), $^{20}$Ne (cyan-blue), 
and $^{24}$Mg (red) in the surface region around the $1.0~M_\sun$ CO WD in  
very late stages of He flashes when the accretion has restarted. 
The surface is toward the left while the WD center is toward the right. 
(a) M10.6. (b) M10.3. (c) M10.16. (d) M10.1. 
The mass-accretion rates are depicted in each panel.  
\label{m10.qx4}}
% source: rnhe.mgsi/m10.6e-7/qx4.wip
\end{figure}

%Fig.3
%\placefigure{m10.Trho}
\begin{figure}
\epsscale{1.15}
\plotone{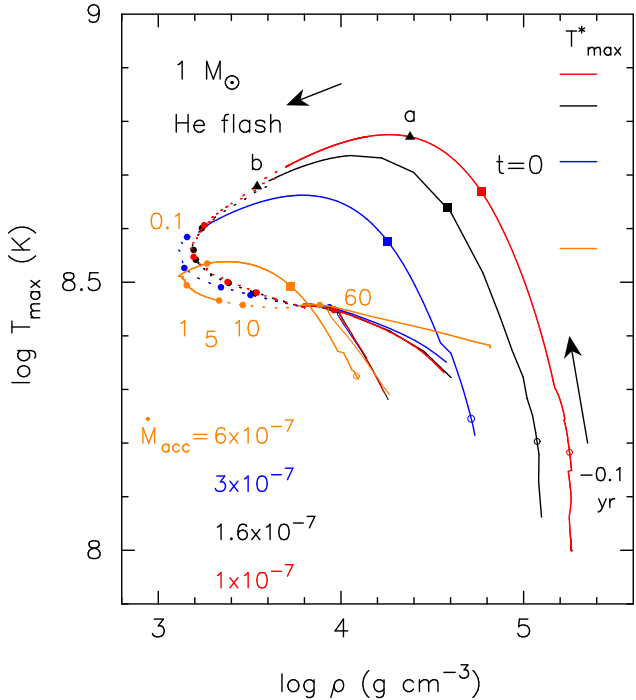}
%%%\plotone{m10.Trho4.f.ps}
\caption{
Locus of the maximum temperature ($T_{\rm max}$) and 
its density for one cycle of our helium flash calculation
on the 1.0 $M_\sun$ WD for various mass-accretion rates. 
The origin (zero) of time is set at the onset of shell flashes 
($L_{\rm nuc}= L_{\rm nuc}^{\rm max}$ at $t=0$).
The indicated time is $t=-0.1$ yr (open circles), 
0.0 (filled squares), 0.1, 1, 5, 10, and 60 yr (five dots, respectively).
Stages labeled a and b (filled black triangles)
correspond to those in Figure \ref{m10.early}. 
The mass loss phase is denoted by the dotted line. After the mass loss phase, 
the nova enters the supersoft X-ray source (SSS) phase
in which the different models 
undergo a similar path until $t \sim 60$ yr. 
Shortly after, the place of maximum temperature shifts inward
($Y=0$ region).
Thus, we indicate the maximum temperature for the whole WD
(thick solid line) as well as the temperature
at the bottom of the He layer ($Y>0$ region, thin solid line). 
The end point of the line is the epoch when the nuclear luminosity
drops to $ L_{\rm nuc}= 100 ~L_\sun$. 
The next outburst occurs in the middle of the helium layer ($Y > 0$),
thus, these cycles do not close in this plot. 
\label{m10.Trho}}
% source: rnhe.mgsi/m10.6e-7/Trho4.f.wip
\end{figure}

%Fig.4
%\placefigure\label{m10.early}
\begin{figure}
\epsscale{1.15}
\plotone{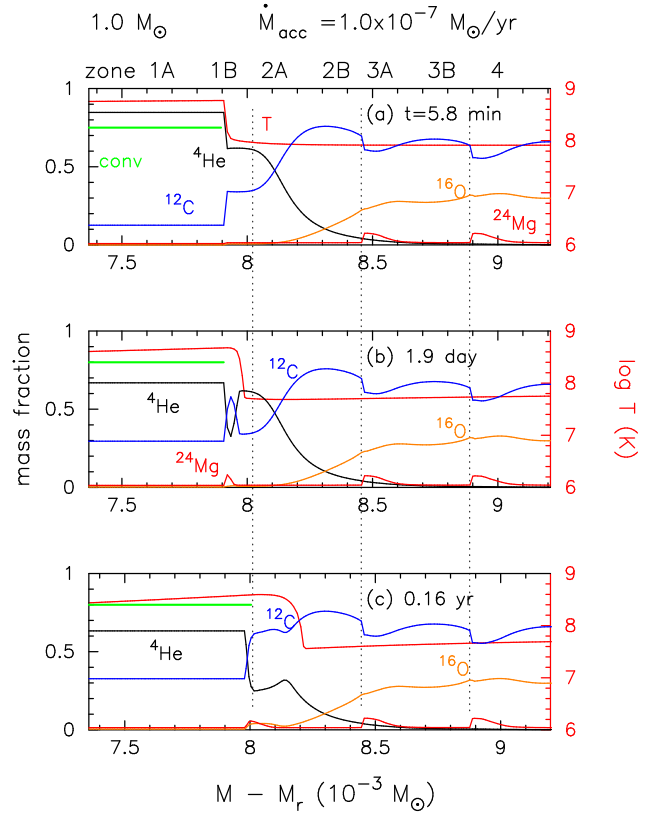}
%\plotone{m10.1e-7.qx.nucbun.early.ps}
\caption{
Distributions of the temperature (upper red solid line), 
$^4$He (black solid), $^{12}$C (blue solid), $^{16}$O (orange solid), 
and $^{24}$Mg (lower red solid line) 
of the surface region for several selected early phases of a He flash 
on the $1.0~M_\sun$ WD.  
The convective region is indicated by the horizontal thick green line.  
\label{m10.early}}
% source: rnhe.mgsi/m10.1.0e-7/qx.nucbun.early.wip
\end{figure}

%Fig.5
%\placefigure\label{m10.late}
\begin{figure}
\epsscale{1.15}
\plotone{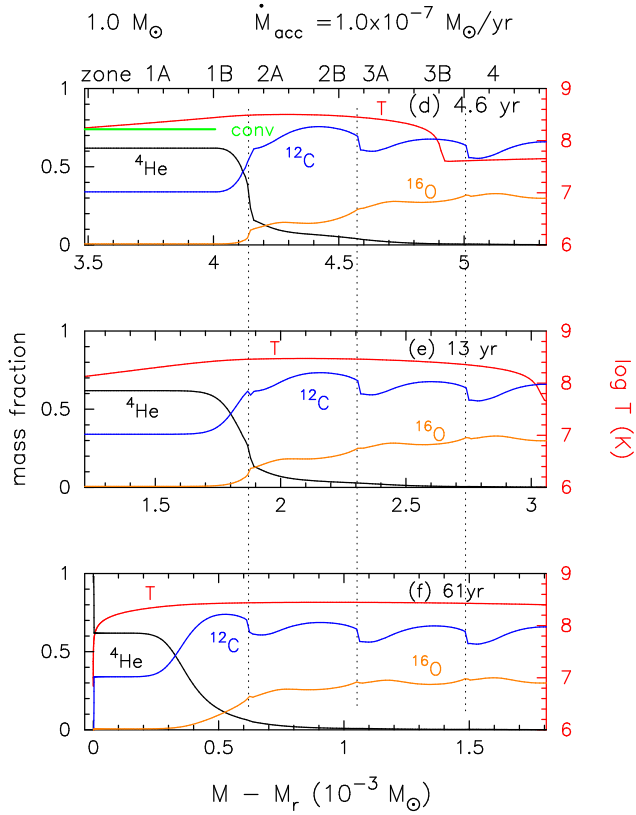}
%\plotone{m10.1e-7.qx.nucbun.late.ps}
\caption{
Same as Figure \ref{m10.early}, but for later phases, and  
$^{24}$Mg is omitted for simplification. 
The mass coordinate is shifted leftward in these stages 
because the envelope mass decreased due to mass loss. 
Mass-accretion restarts in stage (f). 
\label{m10.late}}
% source: rnhe.mgsi/m10.1.0e-7/qx.nucbun.late.wip
\end{figure}

%Table 2
%\placetable{table_hvf}
\begin{deluxetable}{llllllll}
\tabletypesize{\scriptsize}
\tablecaption{Mass Fractions of Nuclear Products}
\tablewidth{0pt}
\tablehead{
\colhead{Model}&
\colhead{$^{12}$C}&
\colhead{$^{16}$O}&
\colhead{$^{20}$Ne}&
\colhead{$^{24}$Mg}&
\colhead{$^{28}$Si$^{a}$}
}
\startdata
M10.6  & 0.66 &0.31 & 0.0029 & 0.0088 & $8.1 \times 10^{-4}$ \\
M10.3  & 0.64 &0.33 & 0.0082    & 0.012 &$8.3 \times 10^{-4}$\\
M10.16 & 0.62 & 0.33 & 0.016 &0.026 & $9.9 \times 10^{-4}$\\
M10.1  & 0.60 & 0.32 & 0.019 &0.035 & $1.5 \times 10^{-3}$\\
M12.6  & 0.59 & 0.24 &0.064 &0.087 & $9.5 \times 10^{-4}$ \\
M12.3  & 0.56 & 0.12 &0.064 &0.24 & $3.6 \times 10^{-3}$\\
M12.16 & 0.53 & 0.087 &0.054 &0.30  & 0.020\\
M135.75  & 0.58  & 0.045 & 0.045 &0.31  &0.0043\\
M135.3  & 0.42 & 0.014 & 0.021 &0.46 &  0.069 \\
M135.16  & 0.33  &0.0087  & 0.014 &0.40  & 0.23 \\
\enddata
\tablenotetext{a}{Including pre-existing Si in the accreted matter $X($Si)=$8.1\times 10^{-4}$.}
\label{table_hvf}
\end{deluxetable}

\section{Results}
\label{section_results}

\subsection{$1.0~M_\sun$ WD}  \label{section_1.0}

Figure \ref{m10.qx4} shows the abundance of various nuclear yields in 
the surface region of the $1.0~M_\sun$ WD with the four different 
mass accretion rates. The photosphere is toward the left 
while the WD center is toward the right. 
There are 41, 43, 25, and 55 small wavy variations in the carbon
and oxygen profiles, corresponding to the temperature variation
during the previous successive helium shell flashes in each model. 
Carbon ($X(^{12}$C) $\sim 0.6$) and oxygen ($X(^{16}$O) $\sim 0.3$) 
are the most abundant elements, which are almost independent of 
the mass accretion rate.  
Helium shell flashes produce small amounts of $^{24}$Mg and  $^{28}$Si
at lower mass-accretion rates. 
The average mass fractions of each element in the wavy profile region 
are summarized in Table \ref{table_hvf}. 

Figure \ref{m10.Trho} shows the temporal changes of the temperature
and density at the point of maximum temperature in the helium-rich
($Y > 0$) region during the last cycle of helium shell flashes
(the 23th cycle for the $1.0\times 10^{-7}~M_\sun$~yr$^{-1}$ case).  
The time from the onset of the flash ($t=0$ defined 
at $L_{\rm nuc}=L_{\rm nuc}^{\rm max}$) is indicated on
the curves in units of year unless otherwise specified. 
The maximum temperature quickly rises in a timescale of 0.1 yr
until $t=0$.  Then, it moves leftward as the envelope begins to expand.
Mass-loss occurs in the dotted part. 
After the maximum expansion, the inner part of the envelope turns
to shrink.  When the photospheric radius shrinks to 
$\log R_{\rm ph}/R_\sun=-1.25$, the mass loss stops. 
Shortly afterward, 
the WD envelope reaches a plane-parallel hydrostatic structure and 
the maximum temperature is almost constant 
($ \log T$ (K) $\sim$ 8.47- 8.5) for several tens of years ($t > 60$ yr). 
This phase corresponds to the supersoft X-ray source (SSS) phase 
($\log T_{\rm ph} \sim 5.5$). 
When the helium-rich envelope mass ($M_{\rm He}=\int d M_r$ for $Y>0$)
decreases and cannot maintain the temperature sufficiently high for He-burning,
the flash ends.  In the post-flash phase, the temperature of envelope
($Y>0$) decreases and the maximum temperature shifts to 
the inner part of the shell ($Y=0$ region).  Therefore, 
in this figure, we plot both the temperature, i.e., 
the maximum temperature (thick solid lines) of the whole WD 
and the temperature (thin solid lines) at the bottom of helium shell.
The next outburst ignites in the mid of the helium-rich region ($Y>0$);
hence, the locus of cycles does not close after one cycle of flashes.  

In Figure \ref{m10.Trho}, a smaller mass-accretion rate model makes
a larger cycle (locus), because the ignition mass is larger
for a low mass-accretion rate, and thus, the temperature 
at the nuclear burning region reaches a higher value.  
The accreted mass $M_{\rm acc}$ before ignition, 
and the maximum temperature $T_{\rm max}^{\rm max}$ for each cycle 
are summarized in Table \ref{table_model}.

Figures \ref{m10.early} and \ref{m10.late} show the temporal change of 
the nuclear products in the very surface region. 
Before the He flash sets in, freshly accreted matter of $Y=0.98$ is 
on top of the leftover of previous outburst 
where the helium mass fraction gradually decreases from $Y=0.62$ to zero. 
Unstable He-burning sets in at the boundary of these two regions 
and the burning region quickly extends upward by convection. 
Figure \ref{m10.early}(a) shows a stage shortly after the onset 
of He flash ($t=5.8$ min). 
The high-temperature region (nuclear burning region) extends
upward (toward the left in Figure \ref{m10.early}) from the boundary. 
As triple-$\alpha$ reaction converts $^4$He into $^{12}$C,
the helium mass fraction decreases with time, 
keeping a constant value with convective mixing 
in Figure \ref{m10.early}(a)--(c) and Figure \ref{m10.late}(d). 
With the progression of time, the high-temperature region gradually extends
inward and reaches zone 2 within 1 yr and zone 3
within 5 yr. 
The leftover helium in zone 2 is consumed by nuclear reactions and 
is exhausted until the end of the shell flash. 
Here, the zone number corresponds to
the cycle number of helium shell flashes from the recent one. 

Figure \ref{m10.early} shows that carbon is synthesized by the 
triple-$\alpha$ reaction in zones 1B and 2A in an early phase of $t <1 $  yr. 
A large part of the carbon produced in zone 1B 
is carried outward by convection to increase the carbon mass fraction 
to $X(^{12}$C$)\gtrsim 0.3$. Correspondingly, the helium fraction
decreases to $Y \sim 0.6$. 
The carbon mass fraction in zone 1B increases after the convection 
disappears, i.e., the envelope becomes radiative,
because the temperature is still as high as $\log T$ (K) $> 8.4$. 
This carbon remains until the next outburst. 

The oxygen mass fraction in zones 1 and 2 increases in the later phase 
(see Figure \ref{m10.late}) through an $\alpha$-capture reaction, 
$^{12}$C$(\alpha, \gamma)^{16}$O,
that continues until the end of the flash. Thus, $^{16}$O 
mass fraction shows wavy variation almost in anti-phase to that 
of $^{12}$C (see also Figure \ref{m10.qx.compari} for a wider mass range
of the envelope). 

Magnesium is produced only at the higher temperature of 
$\log T_{\rm max}$ (K) $> 8.55$, thus, it occurred only in an early phase. 
Figure \ref{m10.early}(b) and (c) show that $^{24}$Mg produced in zone 1B
is carried outward by convection. Only magnesium formed in zone 2A 
(the radiative region) remains after the end of the outburst.

%Fig.6
%\placefigure{m12.Trho}
\begin{figure}
\epsscale{1.15}
\plotone{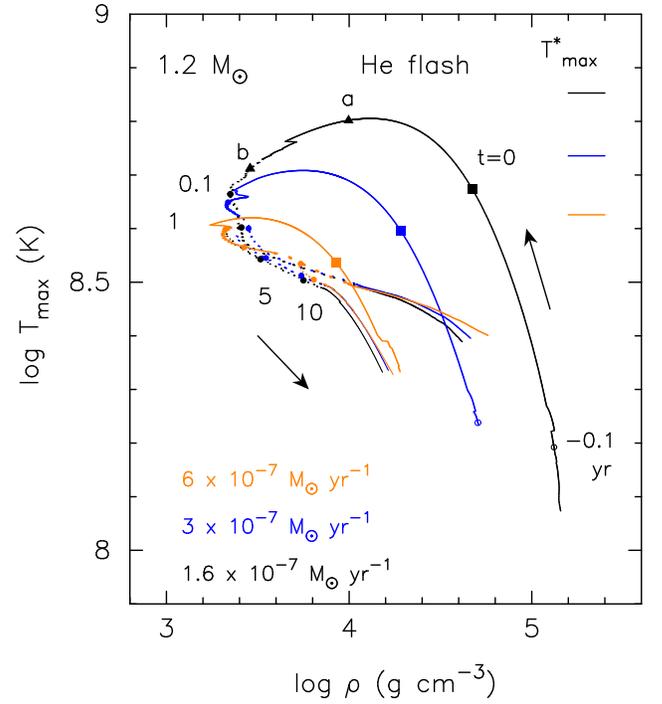}
%\plotone{m12.Trho.ps}
\caption{
Same as Figure \ref{m10.Trho}, but for the 1.2 $M_\sun$ WDs with
the three mass-accretion rates.  
In model M12.16 (black line), the filled black triangles labeled a and b 
correspond to the stages (a) and (b), 
respectively, in Figure \ref{m12qx.early}. 
\label{m12.Trho}}
% source: rnhe.mgsi/m12.1.6e-7/Trho.wip
\end{figure}

%Fig.7
%\placefigure{m12.qx}
\begin{figure}
\epsscale{1.15}
\plotone{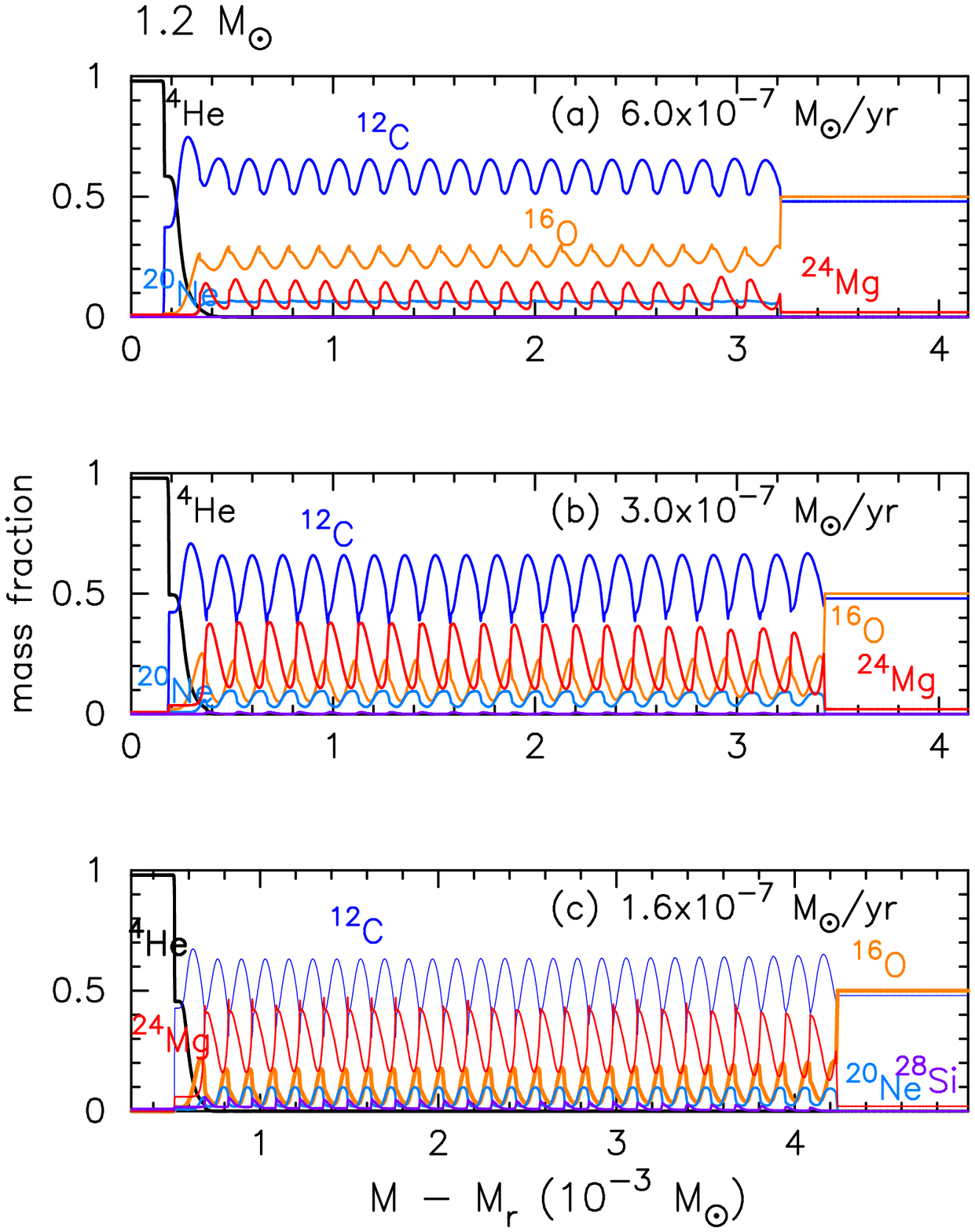}
%\plotone{m12.qx.ps}
\caption{
Same as Figure \ref{m10.qx4}, but for the $1.2~M_\sun$ WD.
\label{m12.qx}}
% source: rnhe.mgsi/m12.1.6e-7/qx.compari.wip
\end{figure}

%Fig.8
%\placefigure{m12qx.early}
\begin{figure}
\epsscale{1.15}
\plotone{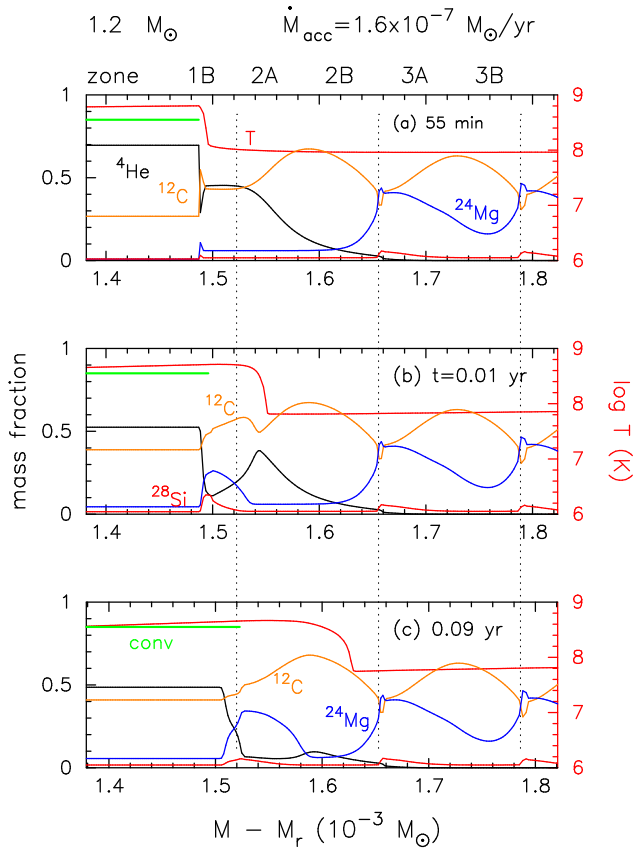}
%\plotone{qx.nucbun.early.ps}
\caption{
Same as Figure \ref{m10.early}, but for the model of M12.16. 
Note that we used different color to Figure \ref{m12.qx} 
to distinguish carbon and He lines: $^4$He (black), $^{12}$C (orange), 
$^{24}$Mg (blue), and $^{28}$Si (lower red).
\label{m12qx.early}}
% source: rnhe.mgsi/m12.1.6e-7/qx.nucbun.early.wip
\end{figure}

%Fig.9
%\placefigure{m12qx.late}
\begin{figure}
\epsscale{1.15}
\plotone{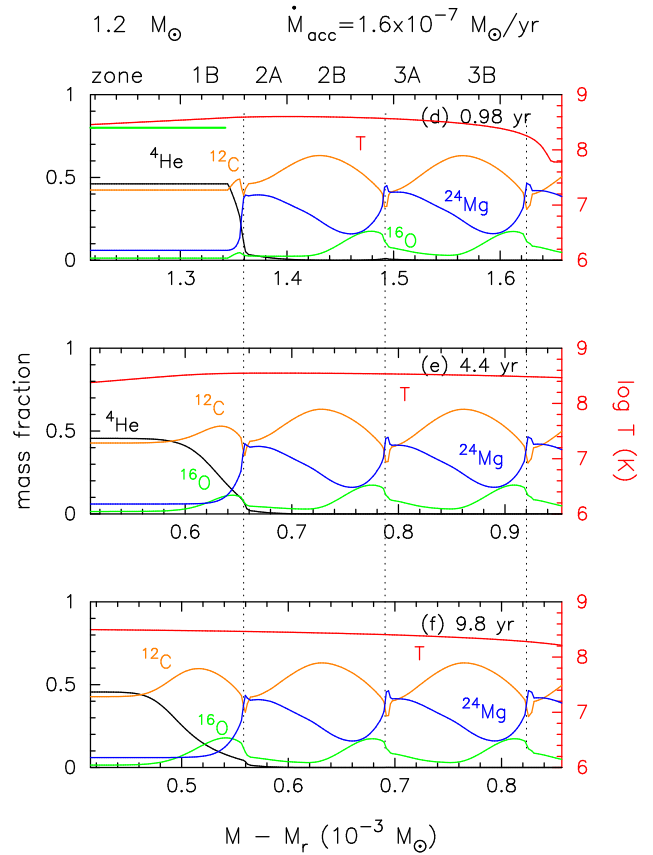}
%\plotone{qx.nucbun.late.ps}
\caption{
Same as Figure \ref{m12qx.early}, but for later phases. 
The abundance profile of $^{16}$O (lower green) is added
and $^{28}$Si is omitted to avoid crowdedness.
\label{m12qx.late}}
% source: rnhe.mgsi/m12.1.6e-7/qx.nucbun.late.wip
\end{figure}

\subsection{$1.2~M_\odot$ WD}\label{section_1.2}

Figure \ref{m12.Trho} shows the temporal change of the maximum
temperature and its density in the He-rich envelope 
during the last cycle of our $1.2~M_\sun$ WD with three different
mass-accretion rates.  The characteristic properties are the same as
those of the $1.0~M_\odot$ WD in Figure \ref{m10.Trho}. 
The maximum temperature is higher than that of the $1.0~M_\odot$ WD
with the same accretion rate, due to a larger gravity, 
despite the smaller ignition mass. 
The flash durations are much shorter than those in the $1.0~M_\odot$ WD, 
but the higher temperature ($\log T_{\rm max}$ (K) $> 8.6$) continues for
a much longer time (1 yr). 
Therefore, the $\alpha$-process produces more massive nuclei. 
Figure \ref{m12.qx} and Table \ref{table_hvf} show $X(^{12}$C)$\sim 0.55$
and $X(^{16}$O) $\sim 0.1 -  0.25$, 
both of which are smaller than those in the $1.0~M_\odot$ WD,
but more $^{24}$Mg is produced. 

Figure \ref{m12qx.early}(b) and (c) show that the $^{24}$Mg and $^{28}$Si produced
in zone 1B are carried out by convection and mixed into the He-rich
envelope. In zone 2A, the radiative zone, the $^{24}$Mg 
fraction increases until $t=1$ yr. 
Only a small amount of $^{28}$Si is produced (see Table \ref{table_hvf}), 
because it is produced by the reaction 
$^{24}$Mg$+\alpha \longrightarrow ~^{28}$Si 
for the high temperatures of $\log T_{\rm max}$ (K) $> 8.7$
which does not last long in the $1.2~M_\sun$ models. 

In stages (e), (f), and after, 
the envelope (zone 1B) becomes radiative.  
The temperature decreases but is still high, $\log T$ (K) $\sim 8.5$, 
so triple-$\alpha$ and $\alpha$-chain reactions, 
$^{12}$C$(\alpha,\gamma)^{16}$O$,(\alpha,\gamma)^{20}$Ne 
produces $^{12}$C, $^{16}$O, and $^{20}$Ne (not plotted) in zone 1B. 
The temperature is not high enough to produce $^{24}$Mg.
Thus, the carbon and magnesium mass fractions show anti-correlation
as shown in Figure \ref{m12.qx}. 
After stage (f), the temperature decreases and the shell flash ends. 
The composition profile in zones 3 and 4 does not change even though
the heat flux passed through inward, 
because there is no He nuclei and $\alpha$-chain reaction does not occur.

%Fig.10
%\placefigure{m135.Trho}
\begin{figure}
\epsscale{1.15}
\plotone{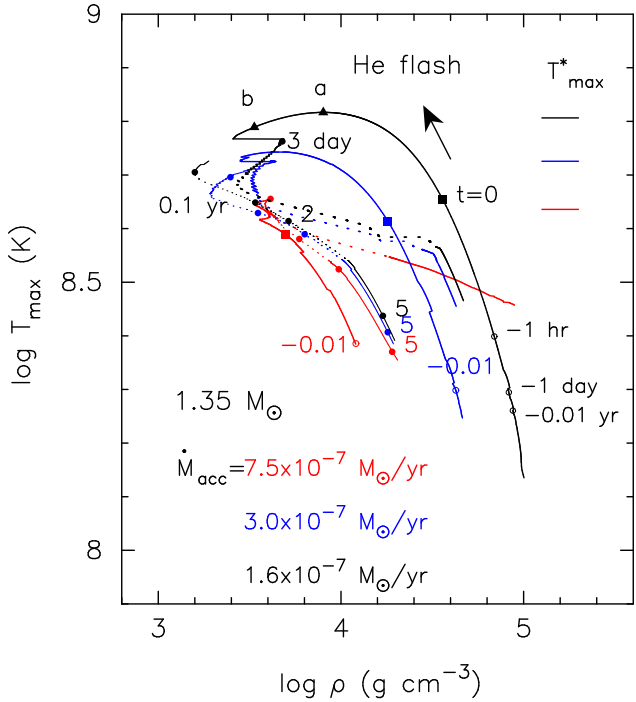}
%\plotone{m135.Trho.ps}
\caption{
Same as Figure \ref{m10.Trho}, but for the 1.35 $M_\odot$ WD.
The quick rightward excursion in M135.16 (black line) at $t=3$ day is due to 
a jump of the place of maximum temperature 
when the He mass fraction at $M-M_r \sim 1.33 \times 10^{-4}~M_\sun$
vanishes ($Y=0$) at $t \sim 3$ day. 
% source: rnhe.mgsi/m135.7.5e-7/Trho.wip
\label{m135.Trho}}
\end{figure}

%Fig.11
%\placefigure{m135qx}
\begin{figure}
\epsscale{1.15}
\plotone{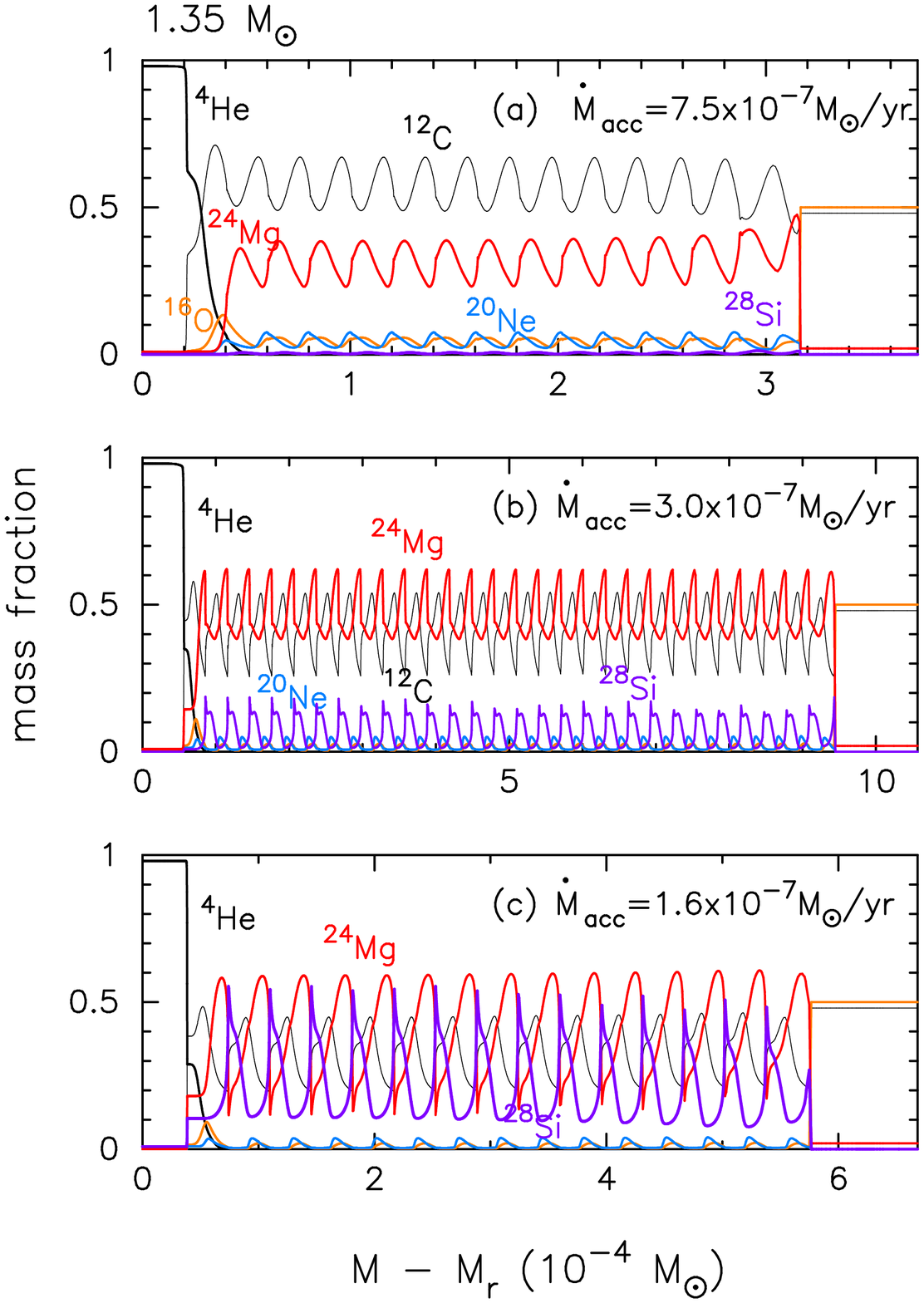}
%\plotone{m135qx.last.ps}
\caption{
Same as Figure \ref{m10.qx4}, but for the $1.35~M_\sun$ WD. 
\label{m135qx}}
% source: rnhe.mgsi/m135.7.5e-7/qx.last.wip
\end{figure}

%Fig.12
%\placefigure
%\placefigure{m135qx.veryearly}
\begin{figure}
\epsscale{1.15}
\plotone{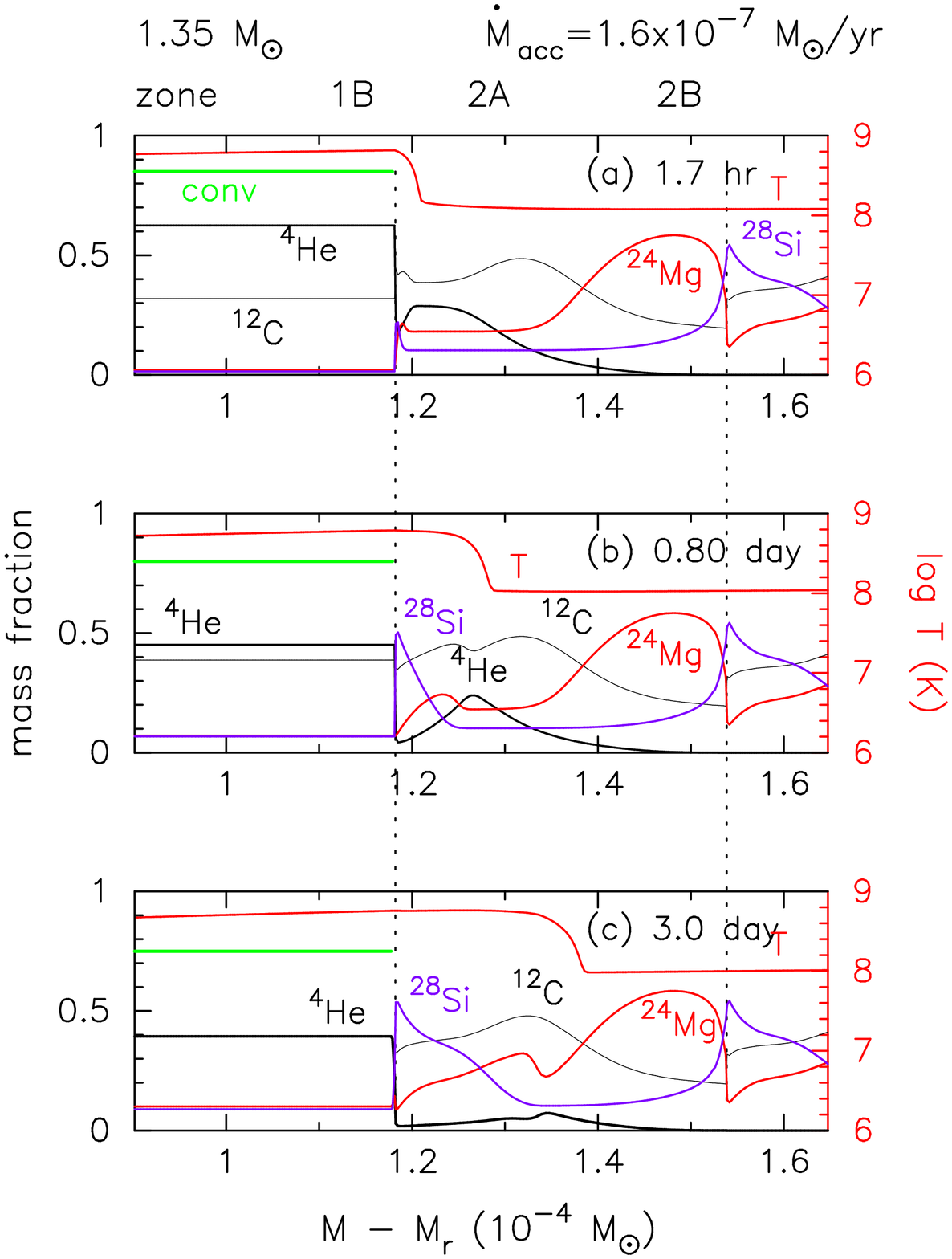}
%\plotone{m135.1.6e-7.qx.nucbun.veryearly.ps}
\caption{
Same as Figure \ref{m10.early}, but for the $1.35~M_\sun$ WD.
\label{m135qx.veryearly}}
% source: rnhe.mgsi/m135.1.6e-7/qx.nucbun.veryearly.wip
\end{figure}

%Fig.13
%\placefigure{m135.qx.Mg}
\begin{figure}
\epsscale{1.15}
\plotone{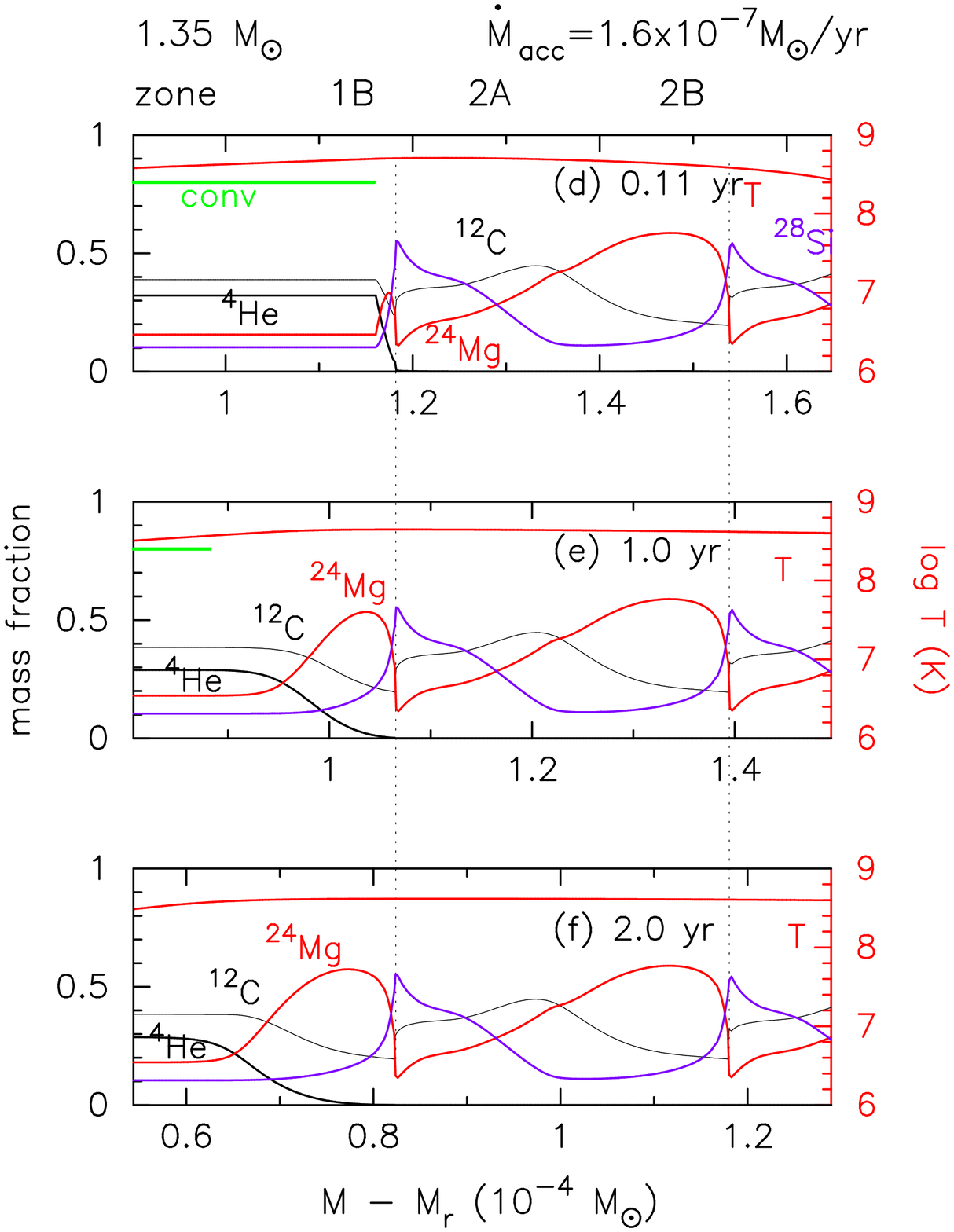}
%\plotone{m135.1.6e-7.qx.nucbun.Mg.ps}
\caption{Same as Figure \ref{m10.late}, but for the $1.35~M_\sun$ WD. 
Mg is produced in zone 1B.
\label{m135.qx.Mg}}
% source: rnhe.mgsi/m135.1.6e-7/qx.nucbun.Mg.wip
\end{figure}

\subsection{ $1.35~M_\odot$ WD}\label{section_1.35}

Figure \ref{m135.Trho} shows the temporal change of $T_{\rm max}$
for the three models of 1.35 $M_\sun$ WD. 
The three models, M135.16, M135.3, and M135.75, follow different paths
in the $\rho$-$T$ plane until $t\sim 1$ yr, 
but similar paths after that.  In model M135.16, the high temperature
($\log T_{\rm max}$ (K) $> 8.6$) period lasts more than two years, 
which is as long as half the total flash duration. 
This longer high-temperature period makes a substantial difference
in nuclear products among the three models, 
as shown in Figure \ref{m135qx}. In model M135.16, $^{24}$Mg 
and $^{28}$Si show much higher peak but less $^{12}$C production.

Figures \ref{m135qx.veryearly} and \ref{m135.qx.Mg} present a close look
at the composition profile in the surface layer 
for the selected stages of model M135.16. The He-shell flash started
at $M-M_r=1.18 \times 10^{-4}~M_\sun$ and 
the high-temperature region widely extends upward with convection. 
In the hot convective region triple-$\alpha$ reaction converts 
$^4$He into $^{12}$C and the $^{12}$C mass fraction increases.
Correspondingly, $^4$He decreases with time. Finally, the composition 
in the convective region consists of $^4$He, $^{12}$C, 
and a few percent of $^{24}$Mg and $^{28}$Si 
(see also Figure \ref{m135qx.ejecta}(b)). 

The hot region extends inward much more slowly with a speed of 
$\lesssim 1 \times 10^{-5}~M_\sun$~day$^{-1}$ as shown in Figure 
\ref{m135qx.veryearly}.  When the temperature rises to $\log T$ (K) $> 8.7$, 
the triple-$\alpha$ reaction and subsequent $\alpha$-chain reactions
up to $^{28}$Si become very active in zone 2A, and as a result, 
$^4$He is completely exhausted. In zone 2, $^{28}$Si is mostly produced 
within 3 days and $^{24}$Mg is produced within one month. 
In the inner zones, $X(^{24}$Mg$)$ and $X(^{28}$Si$)$ do not change 
after the temperature rises, because there is no $^4$He in this region.
As $^{28}$Si is produced from $^{24}$Mg by an $\alpha$-chain reaction,  
$^{24}$Mg decreases when $^{28}$Si increases. 
This can be seen in the periodic anti-phase changes of $^{28}$Si and $^{24}$Mg  
in Figure \ref{m135qx}. 
The mass fraction of $^{24}$Mg in zone 1B increases after the region becomes 
radiative (see Figure \ref{m135.qx.Mg}), but $^{28}$Si does not much increase.
We will compare our results with other calculations 
including large nuclear networks in Section \ref{section_discussion}.

\section{Helium Shell Flashes as a Production Site of HVF\lowercase{s} }
\label{section_production_site}

\subsection{Production of Si}

Table \ref{table_hvf} summarizes the mean values of 
mass fractions of the main nuclei in the wavy composition profiles of
Figures \ref{m10.qx4}, \ref{m12.qx}, and \ref{m135qx}. 
For the $1.0~M_\sun$ WD, the major products of He-burning are C and O,
a few percent of Ne and Mg, and very small amount of Si 
($<$ several $\times 10^{-4}$), independent of the mass-accretion rate.
For the $1.2~M_\sun$ WD, $^{24}$Mg and $^{28}$Si production rates are
very sensitive to the mass-accretion rate.  
For the $1.35~M_\sun$ WD, $^{12}$C and $^{24}$Mg are the most abundant nuclei 
and the $^{28}$Si mass fraction is sensitive to the mass accretion rate. 
The silicon mass fraction increases 
with the decreasing mass-accretion rate, i.e., 
$X(^{28}$Si)$ =3.5 \times 10^{-3}$ for $\dot M_{\rm acc}=7.5 \times 
10^{-7}~M_\sun$~yr$^{-1}$, $0.068$ for $3 \times 10^{-7}~M_\sun$~yr$^{-1}$,
and 0.23 for $1.6 \times 10^{-7}~M_\sun$~yr$^{-1}$. 
(Note that the Si mass fraction in Table \ref{table_hvf} is the summation 
of the newly produced silicon and the pre-existed one in the accreted matter.) 
Thus, we expect more Si could be synthesized 
with much smaller mass-accretion rates. 

Our reaction network does not include nuclei above Si. 
We did not find any other (multi-zone) evolution calculations
involving up to Ca except for the one-zone calculations of He shell flashes.
We will discuss Si and Ca production in more detail 
in Sections \ref{section_nomixing} and \ref{discussion_H}.

\subsection{Stratified surface envelope of WDs and HVFs}

In binary evolution, mass-increasing WDs experience He-shell burning either
when the accreted matter is hydrogen-rich or helium-rich.
In the He star channel to SNe Ia \citep[e.g.,][]{wan17,wu17}, 
a CO WD accretes He matter from a helium star companion and increases
its mass from a lower mass of $M_{\rm WD} \lesssim 1.06~M_\sun$ 
to a mass close to the Chandrasekhar mass.  During the mass increasing stage, 
the WD experiences steady He-burning or periodic He shell flashes, 
depending on the He mass-accretion rate.  The He ash piles up on the CO core. 
The composition of the He ash depends on the WD mass and the mass
accretion rate. 
Generally, the mass transfer rate gradually decreases as the WD mass
increases with time.  This is because the companion mass gradually
decreases due to mass transfer.
Thus, we can expect that the Si-rich layer develops when the WD is
as massive as or more massive than $\sim 1.35~M_\sun$, 
and a Ca-rich layer in the envelope could develop when the WD mass approaches
the Chandrasekhar mass, e.g., $\sim 1.38~M_\sun$.  In this way, 
the WD develops a stratified surface layer composed of different nuclei.
Typically, the most inner stratified layer, just above the CO core,  
is mainly composed of C and O because it was the He ash 
when the WD mass was $\sim 1.0~M_\sun$.  This layer is surrounded 
by the Si-rich layer that is formed when the WD mass was 
$ 1.2-1.35~M_\sun$.  The outermost layer could be Ca-rich 
when the WD grows to as massive as $\sim 1.38~M_\sun$ 
and the mass-accretion rate has decreased significantly. 
%(Note that there is a small amount of Ca ($X$(Ca)$=7.5 \times 10^{-5}$) 
%which is contained in the original accreted matter and remains unprocessed.) 

When the WD finally explodes as an SN Ia, the detonation/deflagration wave
goes through the WD and blows up the whole WD. 
The blast wave activates nuclear reactions and synthesizes heavy 
nuclei depending on its temperature and density.
When the blast wave reaches the surface,
however, the temperature is reduced and 
is not high enough to activate nuclear reactions. 
Thus, the surface region remains unburnt, but is accelerated to high
velocities of $\sim 23000$~km~s$^{-1}$ \citep[e.g., see][]{ger04}. 
In this stratified envelope, the outermost Ca-rich layer receives
the highest expansion velocity and the slightly inner Si-rich layer 
gets a smaller velocity. 
These surface regions could be a possible source of HVFs.

This stratification layer naturally explains the observed 
properties of HVFs as follows:

\noindent
(1) Velocities of \ion{Ca}{2} NIR triplet HVFs are larger 
by $\sim 4000$~km~s$^{-1}$ than that of \ion{Si}{2} $\lambda 6355$
\citep{chi14, zha15, zha16}.  This can be explained by the stratified
layer by our proposed scenario that He shell flashes produce Si-rich layer 
earlier than the Ca-rich layer.  Thus, calcium mainly distributes in the
outer layer than that of Si and, as a result, is accelerated to
larger velocities. 

\noindent
(2) Some SNe~Ia also show \ion{O}{1} $\lambda 7773$ HVFs and its line 
strength is inversely correlated with that of \ion{Si}{2} $\lambda 6355$
(or \ion{Ca}{2} NIR triplet) \citep{zha16}.  
This tendency is consistent with our results 
as listed in Table \ref{table_hvf}.

%Fig.14
%\placefigure{m10.qx4.ejecta}
\begin{figure}
\epsscale{1.15}
\plotone{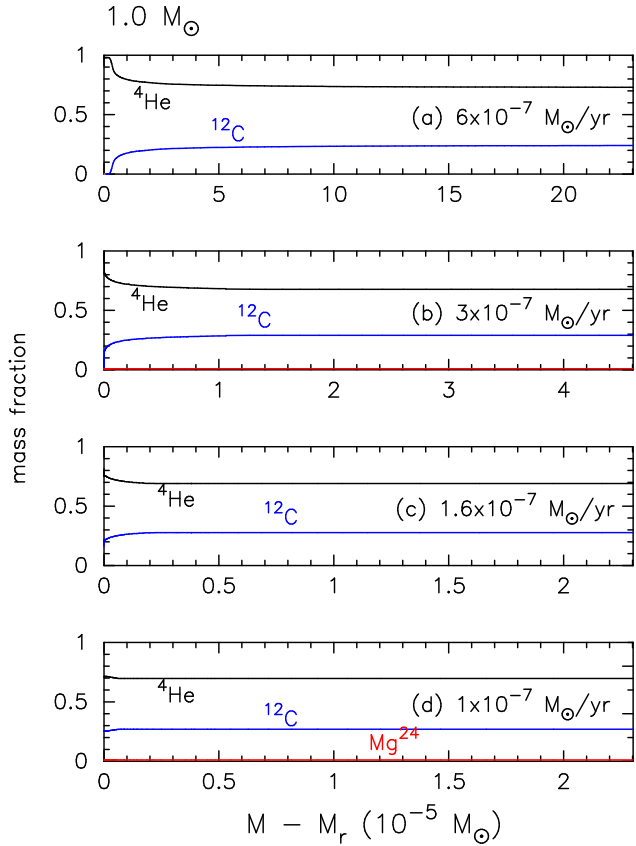}
%\plotone{m10qx.ejecta.ps}
\caption{
Mass fraction of major nuclei in the surface zone of the helium 
envelope just before the mass-loss sets in. The WD mass is $1.0~M_\sun$
with different mass-accretion rates. 
(a) M10.6 at t=4.8 yr. (b) M10.3 at $t=23$ day. (c) M10.16 at $t=1.7$ day.
(d) M10.1 at $t=8.9$ hr.   The surface layer of 
$Y= 0.98$ is very thin hence it cannot be seen in this figure except (a).
In the wind phase, a part of this surface layer will be blown out. 
\label{m10.qx4.ejecta}}
% source: rnhe.mgsi/m10.6e-7/qx4.ejecta.wip
\end{figure}

%Fig.15
%\placefigure{m12qx.ejecta}
\begin{figure}
\epsscale{1.15}
\plotone{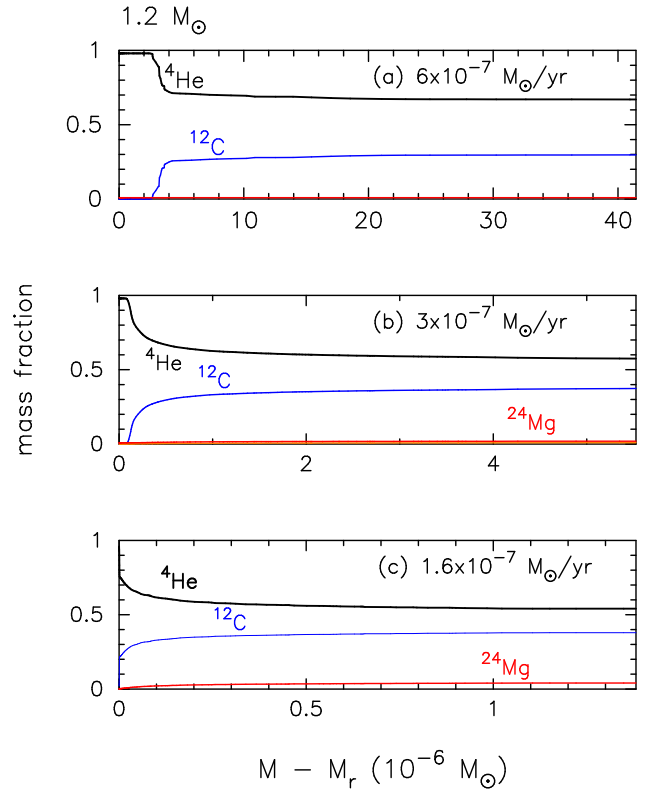}
%\plotone{qx.m12.compari.ejecta.ps}
\caption{
Same as Figure \ref{m10.qx4.ejecta}, but for the $1.2~M_\sun$ WD.
(a) M12.6 at $t=2.3$ yr. (b) M12.3 at $t=0.4$ yr. 
(c) M12.16 at $t=1.9$ day.  
\label{m12qx.ejecta}}
% source: rnhe.mgsi/m12.1.6e-7/qx.compari.ejecta.wip
\end{figure}

%Fig.16
%\placefigure{m135qx.ejecta}
\begin{figure}
\epsscale{1.15}
\plotone{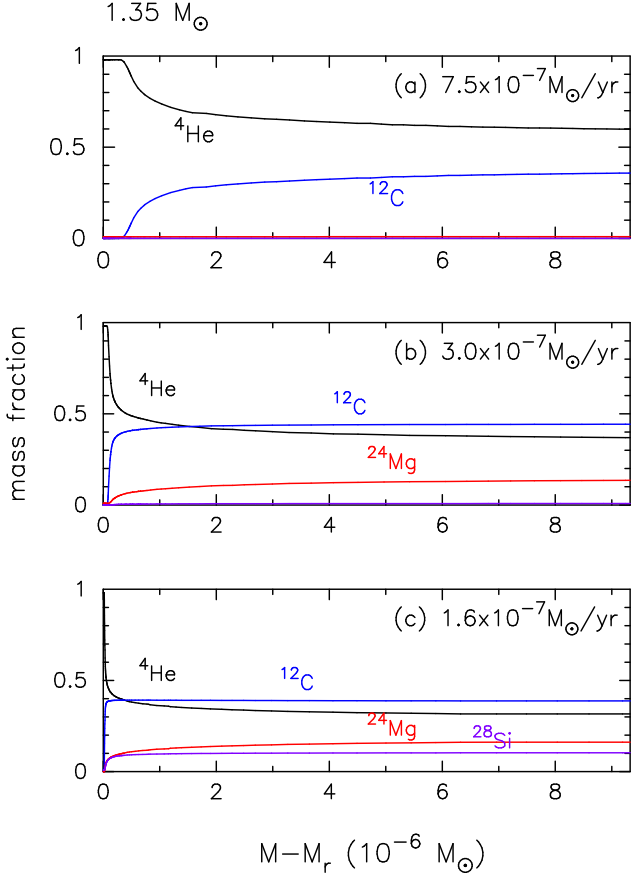}
%\plotone{m135.qx.ejecta.ps}
\caption{
Same as Figure \ref{m10.qx4.ejecta}, but for the 1.35 $M_\sun$ WD. 
(a) M135.75 at $t=0.46$ yr, (b) M135.3 at $t=0.29$ yr, 
(c) M135.16 at $t=0.14$ yr.   
\label{m135qx.ejecta}}
% source: rnhe.mgsi/m135.1.6e-7/qx.ejecta.wip
\end{figure}

\section{Ejecta Composition in He Novae}
\label{section_he_novae}

During a He nova outburst, a part of the envelope is blown off in the wind. 
Figures \ref{m10.qx4.ejecta}, \ref{m12qx.ejecta}, and \ref{m135qx.ejecta}
show the mass fractions of major nuclei just before mass loss occurs.  
The surface region with $Y=0.98$ is made of freshly accreted matter
and the region underneath is the convectively mixed part during the flash. 
At first, the very surface layer of $Y=0.98$ is blown off. 
As time goes on, the inner part will be ejected.  Thus, the composition 
of the ejecta changes with time.  From these figures, we can say that
the envelope consists mainly of He and C, and slightly 
contaminated by Mg and Si, which depends on the WD mass.  

Such a He-flash was realized as the He nova V445 Pup \citep{kat03},
which was discovered on UT 2000 December 30 by Kanatsu \citep{kan00}. 
Unfortunately, the strong dust blackout occurred 210 days after the
discovery.  The outburst shows unique properties such as the
absence of hydrogen and unusually strong carbon emission lines as well as
strong emission lines of Na, Fe, Ti, Cr, Si, Mg, and He 
\citep{ash03, iij08}.  \citet{kat08} estimated the WD mass to be extremely
large ($> 1.35~M_\sun$) using model light curve fitting. 
No calculation results on the abundance ratio were published, 
but the C-rich spectra with He, Si, and Mg lines are broadly consistent
with our model of M135.16.

%Fig.17
%\placefigure{m10.Trho.compari}
\begin{figure}
\epsscale{1.15}
\plotone{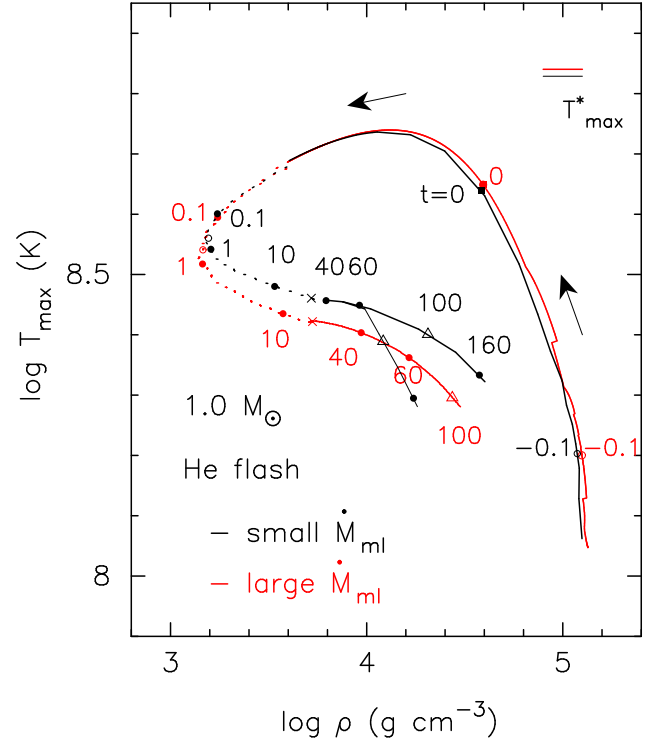}
%\plotone{m10.Trho.compari.ps}
\caption{
Same as Figure \ref{m10.Trho}, but for two models for the 1.0 $M_\sun$ WD
with the same mass-accretion rate but different mass loss rates; 
M10.16 (black line) and M10.16.T (red line). 
Time from $t=0.0$ (at $L_{\rm nuc}= L_{\rm nuc}^{\rm max}$) is indicated
beside the lines in units of years: 
$t=-0.1$ yr (open circle), 0.0 (filled square), 0.1, 
0.5 (open circle), 1, 10, 20 (cross), 40, 60, 100 (open triangle),
and 160 yr (M10.16 only). 
\label{m10.Trho.compari}}
% source: rnhe.mgsi/m10.opal1.6e-7/Trho.wip
\end{figure}

%Fig.18
%\placefigure{m10.qx.compari}
\begin{figure}
\epsscale{1.15}
\plotone{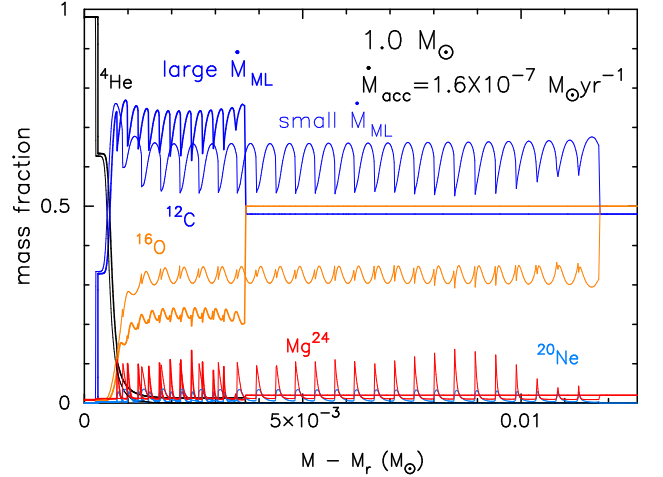}
%\plotone{m10.qx.opaccomp.ps}
\caption{
Comparison of the abundance profiles in the surface region 
of the $1.0 ~M_\sun$ WD models. 
Thin lines: model M10.16. 
Thick lines: M10.16.T the model with extremely large mass-loss rates. 
\label{m10.qx.compari}}
% source: rnhe.mgsi/m10.opal1.6e-7/qx.comp.wip
\end{figure}

\section{Discussion}
\label{section_discussion}

\subsection{Numerical mass-loss rates}

In the present paper, we assumed a mass-loss formula in the extended phase
of the nova outburst as described in Section \ref{section_method}. 
This mass-loss formula may not be accurate because we did not take 
the fitting method as described in \citet{kat17sha} into consideration. 
We call our assumed mass-loss rate the numerical mass-loss rate to
distinguish it from the real wind mass-loss rate.  Here we discuss
the effects of the numerical mass-loss rate on the yield of nuclear synthesis. 

A number of nova light curves have been reproduced theoretically
with optically thick winds 
\citep[e.g.,][]{hac06kb, hac10k, hac14k, hac15k, hac16k, hac16kb}. 
The light curves are well explained in terms of free-free emission,
which is calculated mainly from the wind mass-loss rate. 
At the optical maximum, the wind mass-loss rate is the largest 
and subsequently decreases with time. 
Thus, the optical light curve decay corresponds to the decreasing 
mass-loss rate and decreasing envelope mass. 

We compare two nova light curves with different mass accretion rates.
In the decay phase of the nova outburst, the outbursting envelope 
approaches steady state. 
Then the two nova light curves are almost the same, except for
the peak brightness.  For example, nova 1 has a more massive 
ignition mass, that is, a larger initial envelope mass, and
therefore, has a larger wind mass-loss rate and brighter peak.
After $\Delta t$ from the peak, the envelope mass of nova 1 
decreases and becomes equal to that of nova 2 at the peak. 
So, we have $\Delta t = \Delta M_{\rm acc}/ \dot M_{\rm wind}$, where 
$\Delta M_{\rm acc}$ is the difference of the accreted mass 
between nova 1 and nova 2 and $\dot M_{\rm wind}$ is 
the mean mass loss rate of nova 1 around the peak. 
As $\dot M_{\rm wind}$ is large (because of the initial phase), 
$\Delta t$ is small. 
Thus, the duration of nova 1 is slightly longer than that of nova 2. 

Table \ref{table_model} summarizes our model parameters, 
in which the nova duration is longer for a larger mass-accretion rate 
except the $1.35~M_\sun$ cases. This tendency is not consistent with
the aforementioned theoretical expectation and indicates that we numerically
assumed a larger mass-loss rate for smaller $\dot M_{\rm acc}$ models.

In order to examine the effects of a larger numerical mass-loss rate, 
we have calculated a test model (M10.16.T, T refers to 'test') with 
an extremely large numerical mass-loss rate that is 36 times larger than the maximum rate
for M10.16.   Table \ref{table_model} lists its model parameters. 
Compared with M10.16, the recurrence period and the accreted mass 
are both increased by 11\%. Accordingly, the shell flash  
becomes stronger but only slightly; $L_{\rm nuc}$ is increased 
only by a factor of 1.3. 
This increase is negligibly small compared with the difference
between other $1.0~M_\sun$ models.    

The locus of the maximum temperature in Figure \ref{m10.Trho.compari}
hardly changes from that of M10.16 in the early phase ($t < 1$ yr). 
The duration of the outburst is much shorter (about a half) 
because we assumed a large mass-loss rate. 
Due to the strong mass loss, the envelope mass quickly decreases 
and the nuclear burning turns off 
much earlier (each line ends at $L_{\rm nuc} = 100 ~L_\sun$). 
In short, the early phase evolution of the flash 
(high-temperature period, $\log T$ (K) $>$ 8.6) 
is hardly affected by the choice of the numerical mass-loss rate, 
but the later phase evolution becomes short for the larger mass-loss rate. 

Such difference in evolution results in different nuclear products. 
Figure \ref{m10.qx.compari} compares the composition profile 
of the two models. 
The mass fraction in M10.16T shows 13 peaks in $^{12}$C 
(12 peaks in other nuclei), corresponding to the 13 outbursts. 
The mass fraction of $^{20}$Ne and $^{24}$Mg 
show similar peaks to that of M10.16, because they are produced
in the early high-temperature period.

On the other hand, $X(^{12}$C) is larger in model M10.16T by $0.1$, 
and $X(^{16}$O) is smaller by the same amount. As explained 
in Section \ref{section_1.0}, $^{16}$O is produced from the reaction 
$^{12}$C$+ \alpha$ in $\log T$ (K) $\sim 8.4-8.5$. 
In model M10.16T the temperature quickly drops, 
so this reaction stops earlier than in M10.16. 
This is the reason that $X(^{20}$Ne) and $X(^{24}$Mg) are almost the same 
but $X(^{12}$C) is larger and $X(^{16}$O) is smaller in model M10.16T.  

Thus, if we could assume a smaller mass-loss rate than 
in Table \ref{table_model} for smaller mass-accretion models, 
we expect $X(^{12}$C) and $X(^{16}$O) to change by a significant amount
but $X(^{20}$Ne) and $X(^{24}$Mg) hardly change. 
Figure \ref{m10.qx4} shows that $X(^{12}$C) and $X(^{16}$O) are
almost the same for different mass accretion rates. 
This suggests that the period of $^{16}$O production 
(i.e., $\log T$ (K) $= 8.45-8.5$) is almost the same 
for different mass loss rates (see Figure \ref{m10.Trho}).
Thus, we can say that our numerical mass loss assumption is not
inappropriate from the viewpoint of nuclear products 
on the 1.0 $M_\sun$ WD. 
So we conclude that for the 1.0 $M_\sun$ WD a small amount of 
$^{24}$Mg is produced, but $^{28}$Si is hardly produced. 

The same argument can be applied to the $1.2~M_\sun$ models. 
In models M12.6 and M12.3, the wind phase is relatively short
and $^{24}$Mg production ($\log T$ (K) $\gtrsim $ 8.6) had almost
finished until the mass loss started.  Thus, our main conclusion is that 
a small amount of $^{24}$Mg is produced and no heavier nuclei production
than $^{24}$Mg, is still valid.

In the case of M12.16, $^{24}$Mg is synthesized mainly in the early stages 
(Figure \ref{m12qx.early}(b) and (c)).  The mass loss starts shortly 
before stage (b), and the wind mass-loss rate increases with time
and reaches a value near maximum in stage (d).  
Thus, the resultant mass fraction of $^{24}$Mg depends on 
the adopted mass-loss rate.  In stage (b), the mass loss rate 
$\dot M_{\rm ML}= 2.5 \times 10^{-6}~M_\sun$~yr$^{-1}$, 
is much smaller than the rate of mass reduction due to nuclear burning, 
which is roughly estimated as 
$\dot M_{\rm nuc}= L_{\rm nuc}/ \epsilon_{\rm nuc}= 
4.7 \times 10^7~L_\sun /10^{18}$erg~s$^{-1}
=2.8\times 10^{-3}~M_\sun$~yr$^{-1}$. 
Here we simply assume $\epsilon_{\rm nuc}
=1.\times 10^{18}$erg~s$^{-1}$, including nuclear 
energy generation of He-burning and a part of the burning of the heavier nuclei. 
In stage (c) the mass loss rate $\dot M_{\rm ML}
= 3.2 \times 10^{-5}~M_\sun$~yr$^{-1}$
is also smaller than the nuclear burning rate 
$\dot M_{\rm nuc}=2.3 \times 10^{-4}~M_\sun$~yr$^{-1}$. 
In stage (d) the nuclear burning rate decreases 
and approaches steady state, and  
$\dot M_{\rm ML}= 2.7 \times 10^{-4}~M_\sun$~yr$^{-1}$ is much larger than 
$\dot M_{\rm nuc}= 9.5 \times 10^{-6}~M_\sun$~yr$^{-1}$. 
Thus, if we assume much larger mass-loss rates than we assumed
in the present work, the high-temperature period becomes shorter,
and less $^{24}$Mg is produced. Inversely, if we assume a smaller
mass-loss rate, the period from (b) to (c) does not change, 
but the period from (c) to (d) becomes longer. Therefore, 
the production of $^{24}$Mg could increase slightly. 

Figure \ref{m12qx.early} shows a small amount of $^{28}$Si is produced
in the stages (b) and (c).  In these stages, the mass-loss rate does not
affect evolution, so the production of $^{28}$Si is unchanged. 
Since $^{28}$Si production is small, we may conclude that heavier nuclei
such as $^{40}$Ca are unlikely to be produced. 

In the $1.35~M_\sun$ models, $^{28}$Si is not produced much
for high mass-accretion rates, because the maximum temperature is 
not sufficiently high. Hence, we do not expect significant production of $^{28}$Si and 
heavier nuclei if we decrease the mass-loss rate. 
In model M135.16, most of $^{28}$Si is synthesized  
before the mass loss starts ($t=3$ days in stage (c) 
in Figure \ref{m135qx.veryearly}). Thus the production of $^{28}$Si
does not change much with the choice of numerical mass loss rate.

\subsection{Comparison of nucleosynthesis with other works}
\label{discussion_nuclei}

\citet{has83} calculated nucleosynthesis during explosive He-burning
with a nuclear reaction network for 181 nuclear species from $^1$H 
through $^{62}$Cu, assuming a constant pressure and radiation is dominant.
The nuclear products are dominated by $^4$He and $^{12}$C 
followed by $^{28}$Si and $^{24}$Mg, 
in the model of $\log P$ (dyn cm$^{-2})=21$ 
at $\log T$ (K) $\sim 8.81$ and $\log \rho$ (g cm$^{-3}$) $\sim 4.0$.   
Heavier nuclei are synthesized with higher temperature ($^{32}$S 
at $\log T > 8.85$) and $^{36}$Ar is produced 0.15 \% 
only at the highest temperature, $\log T=8.887$. 
No other heavier nuclei are synthesized.
Although their model is based on the so-called one-zone model,
these results are consistent with the nuclear products of our M135.16 model.
In their model of $\log P$ (dyn cm$^{-2})=22$, $^{40}$Ca is produced at 
a very high temperature $\log T$ (K) $> 9.11$, which is not 
reached in our models (see $T_{\rm max}$ in Table \ref{table_model}). 
We conclude that $^{40}$Ca is hardly synthesized in our $1.35~M_\sun$ WD
models.

\citet{kam12} calculated nucleosynthesis in He shell flashes on massive WDs 
with the so-called one-zone approximation. 
For his 1.35 $M_\sun$ WD model with the envelope mass of  
$3.2 \times 10^{-4}~M_\sun$, the envelope composition 
is $Y= 0.42$, $X(^{12}$C)$= 0.15$, 
$X(^{16}$O)$= 7.8\times 10^{-4}$, $X(^{20}$Ne)$= 0.0025$, 
$X(^{24}$Mg)$= 0.032$, $X(^{28}$Si)$= 0.30$, $X(^{32}$S)$= 4.1\times 10^{-5}$, 
and $X(^{36}$Ar)$= 5.3\times 10^{-4}$ by mass. 
The envelope mass ($3.2 \times 10^{-4}~M_\sun$) is 
about 2 times and the maximum temperature 
($\log T_{\rm max}$ (K) =8.92) is much higher 
than our evolution model M135.16. 
Thus, the nuclear reaction proceeds up to much heavier elements. 
In Kamiya's model, the most abundant nuclei are
$^{28}$Si, followed by $^{12}$C, except unburnt helium. This is consistent 
with the tendency of decreasing (increasing) $^{12}$C ($^{28}$Si) fraction 
extrapolated from our Models M135.75 and M135.16. 
Also, the small amounts of $^{16}$O and $^{20}$Ne are similar to ours. 
Thus, we regard that our nucleosynthesis is consistent with Kamiya's results. 

\citet{kam12} also showed that $^{40}$Ca is hardly synthesized in the 
WD masses of $ < 1.3~M_\sun$.  For his $1.35~M_\sun$ WD model,
a very small amount ($\sim 10^{-5}$) of $^{40}$Ca is produced
for the massive envelope of mass $10^{-3}~M_\sun$.  
We cannot directly connect this one-zone model with our calculation,
but we may say that a significant amount of $^{40}$Ca cannot be
synthesized even if we extend our nuclear network to Ca.

\citet{wu17} presented long-term evolutions of He-accreting WDs
calculated with MESA code including larger nuclear network than ours.  
Their Figure 6 shows the composition profile 
when the initial 1.0 $M_\sun$ WD had grown up to 1.378 $M_\sun$
and the central carbon burning had just begun 
after successive He shell flashes 
with $\dot M_{\rm acc}=7.5 \times 10^{-7}~M_\sun$~yr$^{-1}$. 
When the WD mass was $\sim 1.35 ~M_\sun$ 
($-1.69$ in their log mass coordinate), 
the chemical composition is dominated by $^{12}$C and $^{24}$Mg, 
with $X(^{12}$C) larger than $X(^{24}$Mg) by several tens of percent. 
The secondary dominant elements are $^{16}$O and $^{20}$Ne, 
that are almost the same value 
and one order of magnitude smaller than C and Mg. 
Our model M135.75 (see Figure \ref{m135qx}(a), 1.35 $M_\sun$ with 
$\dot M_{\rm acc}=7.5 \times 10^{-7}~M_\sun$~yr$^{-1}$) shows 
that $X(^{12}$C) varies around 0.6, $X(^{24}$Mg) does around 0.3, 
and both $X(^{16}$O) and $X(^{20}$Ne) are several percent. 
Thus, their results are very consistent with ours.  
\citet{wu17} included 57 reactions up to $^{28}$Si, 
whereas we include 21 reactions also up to $^{28}$Si, 
but both results are quite consistent with each other. Thus, we think 
that our nuclear reaction network covers the major processes
for He flashes.

\citet{wan17} presented He-accreting CO WD models. 
In their pre-central-carbon-ignition model of $1.376~M_\sun$ with 
$\dot M_{\rm acc}=2 \times 10^{-6}~M_\sun$~yr$^{-1}$, 
the envelope is enriched by C and O, i.e., $X(^{12}$C$)\sim 0.45$ and 
$X(^{16}$O$)\sim 0.4$, with little contamination of 
Ne (several percent), Mg, Si ($ < 0.01$), and small amount of 
S ($< 0.001)$. This C/O enrichment for high mass-accretion rate
is consistent with our results.

\subsection{Comparison with other old calculations}

\citet{jos93} calculated He shell flashes with one-zone approximation, i.e. 
plane-parallel structure with no heat flux between the core 
and the He layer, using the Kramers opacity. 
They presented the $\rho$ - $T$ loci for the $1.2~M_\sun$ model, 
similar to our Figure \ref{m12.Trho}. 
Their loci show similar shapes, but located towards the left and
the lower side to our $\rho$ - $T$ loci.  
This difference can be understood from their approximation. 
They assumed that no energy is absorbed in the lower layer 
of the nuclear burning region. 
This approximation is not good \citep[see ][]{kat17sha}, 
and tends to weaken the flash. 
Thus, they obtained lower maximum temperature.

\citet{sha94} presented He flash calculations with the old opacity 
and claimed that Ne/Mg-rich matter accumulates on a $1.25~M_\sun$ WD. 
In their model of $1.25~M_\sun$ WD with 
$\dot M_{\rm acc}=1 \times 10^{-6}~M_\sun$~yr$^{-1}$, 
the ignition mass is $\sim 1 \times 10^{-4} ~M_\sun$ 
and the recurrence period is 100 yr. 
Their Figure 3 shows that $X(^{12}$C) and $X(^{24}$Mg) change with 
anti-correlation around 0.55 and $0.2-0.4$, respectively. 
This anti-correlation and carbon mean fraction is consistent with our 
tendency in Figure \ref{m12.qx}(a).  However, their extremely 
large $^{24}$Mg fraction ($X(^{24}$Mg)=0.55 in the last maximum)
is not consistent with the calculations by \citet{wu17} 
and the present work ($X(^{24}$Mg) $< 0.15$ in model M12.6). 

This overproduction of $^{24}$Mg could be a result of the high maximum
temperature, $\log T_{\rm max}$ (K)=8.72 (see their Figure 2), which 
seems to be extremely high compared with the tendency of our models 
in Table \ref{table_model}. 
Thus, their overproduction could be attributed to 
inadequately larger mass grids for the nuclear burning region, that is too
coarse to resolve the helium-burning layer. 
Such mass grids that are too large are suggested in their Figure 3 that shows only 
five peaks in the $^{12}$C and $^{24}$Mg abundances even after the WD 
experienced eleven shell flashes (see their Figure 2).  
This is due to the rezoning process where they combined neighboring
several mass grids into one to reduce the total number of mass grids
(their referred paper adopted only $< 120$ mass grids for the total WD).
As a result, the abundance profile shows irregular wavy variation. 
The last peak of $X(^{24}$Mg)$\sim 0.55$ is 
3.7 times the previous peak $X(^{24}$Mg)$= 0.15$. 
A small number of mass grids ($< $ a few hundred) is 
insufficient to follow many shell flashes 
because it should cover rapidly changing nuclear burning region
and the expanding outer region    
(see \citet{kat17palermo} for a criticism on their calculation). 
The very high $T_{\rm max}$ could be explained 
as a result of coarse grids in the burning layer. 
We conclude that their claim of a large mass fraction of 
$^{24}$Mg in the $1.25~M_\sun$ WD 
is not real, but a result of too coarse grids calculation.

\subsection{Stability line of He-burning} \label{section_stability}

The WD mass on the stability line for an accretion rate is determined as 
follows. If accretion is started onto a sufficiently less massive WD
(in a left side of Figure \ref{HenomotoD}),  helium burns stably. 
The WD mass increases and moves toward the right in Figure \ref{HenomotoD}.
After a certain mass is accumulated, helium flashes start 
to occur.  We consider the WD mass when the first flash occurred as the mass
of the stability line for the assumed mass accretion rate. 

Figure \ref{HenomotoD} shows the stability line (dashed line) for 
He shell-burning. 
We added another stability line (thin purple solid line)
taken from Figure 1 of \citet{wan17}.
This line agrees well with our stability line 
at $M_{\rm WD} \gtrsim 0.9~M_\sun$. 
The difference of the stability lines in lower mass WDs 
is probably due to the difference
in the stability criterion of different numerical codes.
Near the boundary shell flashes are very weak and difficult 
to draw a clear definition of stability line. 
In our calculation, we regard an evolution with 
small amplitude of luminosity pulsations as ``stable,'' 
because no shell flashes are triggered. 
Although there is no detailed description in \citet{wan17},
we suppose that such difference in the stability definition
is a possible reason for the two different stability lines.

\subsection{Maximum temperature}\label{section_maxT}

The maximum temperature attained in helium shell flashes is important
for nuclear yields.  In this subsection, we examine the constraints
on the maximum temperature.  For a given ignition mass, 
the theoretical maximum temperature can be obtained as follows. 
For a plane parallel envelope, the pressure at the bottom can be 
obtained from a simple integral from the hydrostatic balance, 
$\delta P/\delta r = G M_{\rm WD} \rho /R^2$,
\begin{equation}
P = {G M_{\rm WD}M_{\rm env} \over {4 \pi R^4}},
\label{equation.P}
\end{equation}
where, $M_{\rm WD}$, $R$, and $M_{\rm env}$, are the WD mass, its radius,
and mass of the He-rich envelope respectively.  
We define the maximum temperature $T^*_{\rm max}$ when $P=P_{\rm rad}$, i.e.,  
neglecting the gas pressure, 
\begin{equation}
{T^*_{\rm max}} = ({G M_{\rm WD}M_{\rm env} \over {4 \pi a R^4}})^{1/4},
\label{equation.T}
\end{equation}
where $T^*_{\rm max}$ is the theoretical maximum temperature 
in the extremely radiation dominant plane-parallel atmosphere. 
This temperature is hardly realized in a hydrostatic WD envelope 
because the envelope begins to expand and its configuration changes
to spherically symmetric and the temperature decreases 
before the temperature reaches $T^*_{\rm max}$. 

Figure \ref{m10.Trho} also shows $T^*_{\rm max}$ for each model of 
$1.0~M_\odot$ WD.  The maximum value $T_{\rm max}^{\rm max}$ is smaller 
than $T^*_{\rm max}$ by $\Delta \log T$ (K) $\lesssim 0.11$ (see also 
Table \ref{table_model}). The maximum temperature $T_{\rm max}^{\rm max}$
is closer to $T^*_{\rm max}$ in the 1.2 $M_\sun$ WD (Figure \ref{m12.Trho}), 
and slightly larger than that in the 1.35 $M_\sun$ WD (Figure \ref{m135.Trho}  
and Table \ref{table_model}). This means that thermonuclear runaway produces 
radiation pressure larger than that of the hydrostatic plane-parallel
structure. Thus, the envelope structure changes from plane-parallel
configuration to spherical and the density decrease in a short time. 
This expansion timescale is about an hour for M135.16 as shown 
in Figure \ref{m135.Trho} (point $a$ corresponds $t=1.7$ hour), which
is much longer than the dynamical timescale 
($t_{\rm dyn} \sim 2 \pi \sqrt{R^3/ GM_{\rm WD}} \sim 1$ s).

For a given WD mass, a smaller mass-accretion rate gives a larger
ignition mass.  If the ignition mass is 10 times larger, 
the maximum temperature $T^*_{\rm max}$  is $10^{1/4}$ times larger, i.e., 
it increases by $\Delta \log T^*_{\rm max}$=0.25. 
In the $1.35~M_\sun$ WD, the maximum temperature $T_{\rm max}^{\rm max}$
is close to $T^*_{\rm max}$ as shown in Figure \ref{m135.Trho}, 
thus the maximum temperature could reach as high as 
$\log T$ (K) $\sim 9.0$ for an ignition mass 10 times larger. 
For such a high temperature and massive envelope, 
calcium is synthesized during He flashes 
\citep[$X(^{40}$Ca) $\gtrsim 10^{-4}$,][]{kam12}. 
If we further assume mixing between freshly accreted matter and
ashes, we expect much more production of calcium, as discussed in the
next section.

\subsection{Effect of mixing in the quiescent phase}
\label{section_nomixing}

In the present work we simply assumed that the freshly accreted 
He matter piled up onto the ashes of previous flashes. 
However, some mixing may occur between the freshly accreted matter and
the ashes during the quiescent phase 
(i.e., between the two successive outbursts) 
owing to some mechanism like 
hydrodynamic instabilities \citep[e.g.,][]{pir15}. 

In our models, the outburst begins 
at the boundary between the accreted matter $X(^4$He) =0.98 and the 
leftover ash, i.e., the He ignition starts 
at $M-M_r=1.18 \times 10^{-4}~M_\sun$ in M135.16 
(Figure \ref{m135qx.veryearly}). 
If the mixing occurs the He-rich matter ($X(^4$He) =0.98) mixes into the ashes 
and the helium mass fraction at $M-M_r > 1.18 \times 10^{-4}~M_\sun$ increase. 
This makes the helium ignition occur more inside than our models, i.e,  
the ignition mass becomes larger. 
As in Section \ref{section_maxT} 
a larger ignition mass enables a flash to reach a higher maximum temperature 
in the nuclear burning region. 
Thus, nuclear reaction could proceed to more massive nuclei. 

\citet{pir15} studied turbulent mixing in He-accreting WDs and showed that 
mixing is greatest at low spin, high accretion rate, and high WD mass. 
For $1.3~M_\sun$ and $\dot M_{\rm acc}=1-10 \times 10^{-7}~M_\sun$~yr$^{-1}$, 
the ashes is mixed with the accreted matter about
0.6-0.8 times the accreted matter by weight. 

In order to see how the mixing affects nuclear production, 
we calculated a test model of $1.35~M_\sun$ with $\dot M_{\rm acc}
=1.6 \times 10^{-7}~M_\sun$yr$^{-1}$, i.e., the same parameters 
as those of model M135.16, but the accreting matter has the chemical
composition of $Y=0.5$, $X$(C)=0.14, $X$(Mg)=0.1679, $X$(Si)=0.1765,
assuming the same amount of ash is uniformly mixed with the accreting
matter. The ignition mass is $M_{\rm env}=1.3 \times 10^{-4}
~M_\sun$, slightly increased from model M135.16. 
The maximum temperature reaches $\log T_{\rm max} =8.83$,
increased by $\Delta \log T=0.015$.
The He-burning ash has the chemical composition of $X(^{12}$C)=0.17,
$X(^{24}$Mg)=0.18, $X(^{28}$Si)=0.63,  i.e., more carbon and magnesium are
processed to silicon than in model M135.16. 
This can be understood from Table \ref{table_hvf}, 
that shows silicon production being 
sensitive to the maximum temperature. 
For Ca production, however, we may not expect much, because it is 
produced in much higher temperature ($\log T$ (K) $> 9.0$). 

We can further consider a case that the accreted matter is not
uniformly mixed with the ash, but is divided into many small blobs
of the original composition ($X(^4$He) =0.98) 
that dig into the deep ash and ignite there.  
In this case, the ignition mass effectively becomes larger. 
If we roughly assume an ignition mass as twice as large, i.e., 
the same amount of ash being mixed with the accreted matter,  
the maximum temperature could be higher 
by $\Delta \log T \sim (\log 2)/4 = 0.075$ 
than that in M135.16, from Equation (\ref{equation.T}).
Then, the maximum temperature may reach $\log T =8.81+0.075 =8.89$.
This temperature corresponds to that of 
a Kamiya's  grid model \citep{kam12},
$M_{\rm WD}= 1.38~M_\sun$ with $\log M_{\rm env}\sim -4.0$, that gives  
$\log X(^{40}$Ca) $\sim -6$. 
Thus, a very small but non-zero amount of Ca could be produced. 

For a $1.38~M_\sun$ WD, we may take a model calculation in \citet{kat17shb}
in which the first helium-shell flash occurs after the 1543 successive 
hydrogen flashes for $\dot M_{\rm acc,H}= 1.6 \times 10^{-7}~M_\sun$~yr$^{-1}$.
The ignition mass is $M_{\rm env,He}=7.5 \times 10^{-5}~M_\sun$ and the 
maximum temperature reaches $\log T_{\rm max} = 8.88$.  If we assume that the
mixing makes the ignition mass 2 or 3 times larger, the maximum temperature
would be close to $\log T_{\rm max} = 8.96$, and 9.0, respectively. 
Kamiya's grid models give $\log X(^{40}$Ca$)= -5$ to $-4$, one order of
magnitude larger than in the $1.35~M_\sun$ case. 
However, the Ca production is very sensitive to the temperature, and then 
we cannot draw a definite conclusion using 
Kamiya's one-zone model and Kato et al.'s first He flash model.

\subsection{Hydrogen-accreting WDs} \label{discussion_H}

In a hydrogen-rich envelope with no mixing of core material,
hydrogen burning does not produce massive nuclei such as Si and Ca
because of lower maximum temperatures \citep[e.g., $\log T_{\rm max}
($K$)= 8.23$ for $1.38~M_\sun$][]{kat17shb}.  Thus, we consider a case
of He flash with the H-rich layer on top of it.  \citet{kat17shb}
calculated successive 1543 H-flashes followed by a first He flash. 
They stopped calculation in the midway of the He flash owing to numerical
difficulties, but showed that the surface hydrogen is mixed into 
the He burning zone by convection. 

\citet{wana99} calculated one zone model with a nuclear network up to Ca. 
For an envelope of initial chemical composition of $X$(H)= 0.424, 
$X$(He)= 0.165, $X$(C)= 0.0176, $X$(O)= 0.222, $X$(Ne)= 0.134, and 
$X$(Mg)= 0.0215 with $\log T_{\rm max}$ (K) = 8.86 (their Figure 3), 
proton capture reactions are active and the final product contains  
$X$(Si)=$10^{-3} - 10^{-2}$ and $X$(Ca)$=10^{-4}$.   
\citet{pol95} also calculated a $1.35~M_\sun$ model with 95 mass zones,   
in which the maximum temperature reaches $\log T_{\rm max}$ (K) = 8.55. 
The nuclear network covers up to Ca.  They assumed the initial
chemical composition of $X$(H)= 0.365, $X$(He)= 0.133, $X$(O)= 0.15,
and $X$(Ne)= 0.249, and $X$(Mg)= 0.10.  The final calcium mass fraction
slightly increases to $X$(Ca)$=1.82 \times 10^{-5}$ 
from the initial value of $1.66 \times 10^{-5}$.   

Combined these calculation with mixing effects of making ignition mass larger, 
we expect that a substantial amount of Ca would be produced in a very
massive WD like $\gtrsim 1.38~M_\sun$.  It is however clear that we need
further calculations in order to obtain definite conclusions on Ca
production.

\section{Conclusions}
\label{section_conclusion}

Our main conclusions are summarized as follows. 

\noindent
1. We present successive helium shell-flash calculations 
with detailed description of nuclear products 
on the 1.0 $M_\sun$, 1.2 $M_\sun$, and 1.35 $M_\sun$ WDs with the
He mass-accretion rates of 
$\dot M_{\rm acc}=1.0 - 7.5 \times 10^{-7}~M_\sun$~yr$^{-1}$.
Massive nuclei such as $^{24}$Mg and $^{28}$Si are produced on  
more massive WDs with lower mass-accretion rates.

\noindent
2. A mass-growing WD develops a surface layer as He-burning ash
on top of the WD core.  This surface layer is enriched with Mg and Si,
depending on the mass accretion rate.  The silicon in this surface
layer is a possible source of \ion{Si}{2} $\lambda 6355$ HVFs 
that is often observed in SN Ia spectra in early phases. 

\noindent
3. In the He-star donor channel to SNe~Ia, in which the WD accretes
helium-rich matter, WDs had developed a surface layer enriched 
by $^{24}$Mg and $^{28}$Si only when it approaches the Chandrasekhar
mass, and possibly a small amount of Ca before the explosion.
These could be origins of \ion{Si}{2} $\lambda 6355$ 
and \ion{Ca}{2} NIR triplet HVFs.

\noindent
4. Ejecta of He shell flashes consist mainly of He and C. 
In the $1.35~M_\sun$ WD, the ejecta have $> 30 \%$ of helium,
 $< 40 \%$ of carbon, and are enriched with Mg and Si by several percent
in relatively low mass-accretion rates.  This composition is broadly
consistent with C-rich spectra of the helium nova V445 Pup.

\noindent

\noindent

\acknowledgments
The authors thank the anonymous referee for useful suggestions and comments 
that improve the manuscript. 
 This research has been supported in part by Grants-in-Aid for
 Scientific Research (15K05026, 16K05289)
 of the Japan Society for the Promotion of Science.

%\clearpage

%Table 1

%%%%%%%%%%%%%%%%%%%%%%%   Table 2  \ref{table_hvf}

%Table 2

% todayfigure

%Fig.1

%\clearpage

%%%%%%%%%%%%%%%%%%     1.0  Mo  %%%%%%%%%%%%%%%%%%%

%Fig.2

%Fig.3
%\clearpage

%Fig.4

%Fig.5

%%%%%%%%%%%%%%%%%%     1.2  Mo  %%%%%%%%%%%%%%%%%%%

% Fig.6

% Fig.7

%Fig.8

%Fig.9

%\clearpage

%%%%%%%%%%%%%%%%%%     1.35  Mo  %%%%%%%%%%%%%%%%%%%

% Fig.10

% Fig.11

% Fig.12

%Fig.13

%\clearpage
%%%%%%%%%%%%%%%%%%%%%%%%%     ejecta %%%%%%%%%%%%%%%%%%%%%%%

% Fig.14

% Fig.15

% Fig.16

%\clearpage 

%%%%%%%%%%     comparison of   1.0 Mo  large/small ML rates   %%%%%%%%%%

% Fig.17

% Fig.18

\end{document}